\documentclass[aps,pre,citeautoscript,showkeys]{revtex4-1} 
\pdfoutput=1
\usepackage{graphicx}  
\usepackage{dcolumn}   
\usepackage{bm}        
\usepackage{amssymb}   
\usepackage{epsfig}
\usepackage{epstopdf}  
\usepackage{subfigure}
\usepackage{tensor}
\usepackage[utf8]{inputenc}
\usepackage{mdframed}
\usepackage{amsmath}
\usepackage{lipsum}
\usepackage{protosem}
\usepackage{wasysym}
\usepackage[colorlinks=true]{hyperref}  
\hypersetup{
    bookmarks=true,         
    unicode=false,          
    pdftoolbar=true,        
    pdfmenubar=true,        
    pdffitwindow=false,     
    pdfstartview={FitH},    
    pdftitle={My title},    
    pdfauthor={Author},     
    pdfsubject={Subject},   
    pdfcreator={Creator},   
    pdfproducer={Producer}, 
    pdfkeywords={keyword1} {key2} {key3}, 
    pdfnewwindow=true,      
    colorlinks=true,       
    linkcolor=magenta, 
    citecolor=blue,        
    filecolor=magenta,      
    urlcolor=cyan           
} 

\newcommand{\labell}[1]{\label{#1}}

\newcommand{\bea}{\begin{eqnarray}}
\newcommand{\eea}{\end{eqnarray}}
\newcommand{\ba}{\begin{eqnarray}}
\newcommand{\ea}{\end{eqnarray}}

\newcommand{\beq}{\begin{equation}}
\newcommand{\eeq}{\end{equation}}
\newcommand{\beqa}{\begin{eqnarray}}
\newcommand{\eeqa}{\end{eqnarray}}
\newcommand{\beqar}{\begin{eqnarray*}}
\newcommand{\eeqar}{\end{eqnarray*}}

\newcommand{\eg}{{\it e.g.,}\ }
\newcommand{\ie}{{\it i.e.,}\ }

\newcommand{\E}{\mathcal{E}}

\newcommand{\req}[1]{(\ref{#1})} 

\begin{document}

\title{Four-dimensional black holes in Einsteinian cubic gravity} 
\author{Pablo Bueno$^{(1,2)}$ and Pablo A. Cano$^{(3)}$}
\affiliation{\vspace{0.1cm}$^{(1)}$Instituut voor Theoretische Fysica, KU Leuven,
Celestijnenlaan 200D, B-3001 Leuven, Belgium\\
$^{(2)}$Institute for Theoretical Physics, University of Amsterdam
 1090 GL Amsterdam, The Netherlands\\
$^{(3)}$Instituto de F\'isica Te\'orica UAM/CSIC,
C/ Nicol\'as Cabrera, 13-15, C.U. Cantoblanco, 28049 Madrid, Spain\vspace{0.1cm}}
\date{\today}

\begin{abstract}
We construct static and spherically symmetric generalizations of the Schwarzschild- and Reissner-Nordstr\"om-(Anti-)de Sitter (RN-(A)dS) black-hole solutions in four-dimensional Einsteinian cubic gravity (ECG). The solutions are characterized by a single function which satisfies a non-linear second-order differential equation. Interestingly, we are able to compute independently the Hawking temperature $T$, the Wald entropy $\mathsf{S}$ and  the Abbott-Deser mass $M$ of the solutions analytically as functions of the horizon radius and the ECG coupling constant $\lambda$. Using these we show that the first law of black-hole mechanics is exactly satisfied. 
Some of the solutions have positive specific heat, which makes them thermodynamically stable, even in the uncharged and asymptotically flat case. Further, we claim that, up to cubic order in curvature, ECG is the most general four-dimensional theory of gravity which allows for non-trivial generalizations of Schwarzschild- and RN-(A)dS characterized by a single function which reduce to the usual Einstein gravity solutions when the corresponding higher-order couplings are set to zero.
\end{abstract}

\maketitle
 
 \section{Introduction and summary of results}
 
In \cite{PabloPablo} we showed that, up to cubic order in curvature, the most general dimension-independent gravity theory constructed from arbitrary contractions of the metric and the Riemann tensor whose linearized spectrum coincides with the one of Einstein gravity can be written as a linear combination of the Lovelock terms \cite{Lovelock1,Lovelock2} plus a new cubic contribution $\mathcal{P}$, defined as
\begin{equation}
\mathcal{P}=12 R_{a\ b}^{\ c \ d}R_{c\ d}^{\ e \ f}R_{e\ f}^{\ a \ b}+R_{ab}^{cd}R_{cd}^{ef}R_{ef}^{ab}-12R_{abcd}R^{ac}R^{bd}+8R_{a}^{b}R_{b}^{c}R_{c}^{a}\, ,
\end{equation}
 which we coined \emph{Einsteinian cubic gravity} (ECG) term. Remarkably, as opposed to the quadratic and cubic Lovelock terms --- which are respectively topological and trivial in that case --- this new term is dynamical in four-dimensions. Hence, the $D=4$ ECG action can be written as
\begin{equation}\label{ECGaction}
S=\frac{1}{16 \pi G}\int_{\mathcal{M}}d^4x\sqrt{|g|}\left[-2\Lambda_0+R- G^2 \lambda   \mathcal{P}\right]\, ,
\end{equation}
where $\Lambda_0$ is the cosmological constant, $G$ is the Newton constant and $\lambda$ is a dimensionless coupling constant which we will assume to be positive throughout the paper, \ie $\lambda\geq 0$.  Also in \cite{PabloPablo}, we anticipated that \req{ECGaction} admits static and spherically-symmetric black-hole-like solutions characterized by a single function $f(r)$, \ie metrics of the form
\begin{equation}
\label{ansatz0}
ds^2= -f( r ) dt^2+\frac{dr^2}{f( r )}+r^2 d\Omega_{(2)}^2\, ,
\end{equation}
where $d\Omega^2_{(2)}=d\theta^2+\sin^2\theta d\phi^2$ is the metric of the round sphere. In this paper we will show that this is indeed the case. In particular, we will construct new black-hole solutions of the ECG theory \req{ECGaction} which generalize the usual Schwarzschild black hole --- and reduce to it as $\lambda\rightarrow 0$.  We will show that for general values of $\lambda$, $f(r)$ is determined by a non-linear second-order differential equation --- see \req{fequation} below. We shall not be able to solve this equation analytically. However, we will be able to compute analytic expressions --- valid for general values of $\lambda$ --- for the Hawking temperature $T$, the Wald entropy $\mathsf{S}$ and the Abbott-Deser mass $M$ of the black hole as functions of the new horizon radius $r_h$. Remarkably enough, we are able to show that our black hole solutions exactly satisfy the first-law of black hole mechanics 
\begin{equation}\label{1st}
dM=T d\mathsf{S}\, .
\end{equation}
We stress that our calculations of $T$, $\mathsf{S}$ and $M$ are independent from each other, so the fact that \req{1st} holds is a quite non-trivial fact. Note that the first law was proven to hold for small perturbations of stationary black hole solutions of general higher-order gravities in \cite{Wald:1993nt}, so this is an important check on our solutions. To the best of our knowledge, the solutions presented here constitute the first examples of four-dimensional generalizations of the Schwarzschild and Reissner-Nordstr\"om black holes to any higher-order gravity which are determined by a single function and for which the first law is proven to  hold analytically --- see the Note added at the end of this section though. 

In the asymptotically flat case, given a value of the mass $M$, all solutions with $\lambda >0$ have horizon radius and Wald entropy $\mathsf{S}$ larger than the Schwarzschild ones. Also, some of them are thermodynamically stable --- \ie they have positive specific heat. Interestingly, the temperature of the new solutions, which is always smaller than Schwarzschild's, is bounded from above by 
\begin{equation}
T_{\rm{max}}=\frac{1}{12\pi G^{1/2}  \lambda^{1/4}}\, ,
\end{equation}
which is reached for $M_{\rm max}=16/27 G^{-1/2}\lambda^{1/4}$, and it vanishes both as $M\rightarrow 0$ and as $M\rightarrow +\infty$. 

These results for uncharged asymptotically flat solutions are extended in section \ref{charged} to incorporate a non-vanishing cosmological constant and electric charge --- \ie we add a Maxwell term to the gravitational action \req{ECGaction}. The corresponding solutions, which are again characterized by a single function, generalize the usual Reissner-Nordstr\"om-(Anti) de Sitter  (RN-(A)dS) black hole and we show that they also satisfy the first law, which in that case reads
\begin{equation}
dM= T d\mathsf{S}+\Phi dq,
\end{equation}
where $\Phi=q/(4\pi r)$ is the electrostatic potential.

The structure of the paper goes as follows. In section \ref{ecgse} we show that \req{ansatz0} is a valid ansatz for ECG, and determine the equation satisfied by $f(r)$. In section \ref{asy}, we focus on the uncharged asymptotically flat case. There, we provide asymptotic expressions for $f(r)$ and show that it can describe a black hole with horizon radius $r_h$. Then, we obtain exact expressions for the mass and the surface gravity as functions of $r_h$ and $\lambda$. We also plot numerically the $f(r)$ of our solution for various values of $\lambda$. In section \ref{them} we compute the Wald entropy and the Hawking temperature of the solution, unveiling some interesting differences with respect to the usual Schwarzschild case --- \eg some solutions possess positive specific heat. Finally, we prove that the first law holds exactly for our solutions. As explained above, these results are extended to the charged asymptotically (A)dS case in section \ref{charged}. We conclude in section \ref{discussion}, where we also argue that ECG is in fact the most general four-dimensional theory of gravity which allows for non-trivial single-function generalizations of the Schwarzschild- and RN-(A)dS solutions which reduce to the usual Einstein gravity  ones when the corresponding higher-order couplings are set to zero. We also explain that, as opposed to the $D=4$ case, five-dimensional ECG does not admit single-function solutions. Appendix \ref{remnants} contains some speculations on the possibility that our solutions could describe long-lived microscopic black hole remnants without the ECG coupling affecting the usual macroscopic physics of general relativity.

\textbf{Note added}: As we were writing this paper, \cite{Hennigar:2016gkm} appeared in the arXiv.
This work contains a major overlap with our results in sections \ref{ecgse}, \ref{asy} and \ref{them}. There are some differences though: in \cite{Hennigar:2016gkm}, the fact that $f(r)$ satisfies a second-order differential equation has been overlooked (it is not obvious from the field equations of the theory that the corresponding third-order equation is a total-derivative). As a consequence, the authors do not find an expression for the mass $M$ as a function of the horizon radius. Instead, they use the first law to determine $M$ using the values of $T$ and $\mathsf{S}$ (which agree with ours). Here, we are able to compute $M$ independently, which allows us to verify that the first law holds. Of course, the final result is the same. On the other hand, the charged case is not considered in \cite{Hennigar:2016gkm}.

\section{Spherically symmetric solutions of ECG}\label{ecgse}

 In this section we will show that \req{ECGaction} admits generalizations of the Schwarzschild-(A)dS black hole characterized by a single function whose expression can be determined by solving a second-order differential equation.
Without loss of generality, we assume the following ansatz for our static and spherically symmetric solution \footnote{The extension to hyperbolic and planar horizons is straightforward.}
\begin{equation}
\label{ansatz1}
ds^2= -N^2( r )f( r ) dt^2+\frac{dr^2}{f( r )}+r^2 d\Omega_{(2)}^2\, ,
\end{equation}
Now, a possible route would entail evaluating the field equations of \req{ECGaction} on the above ansatz and finding the corresponding equations for $N(r)$ and $f(r)$. Here we will use a different method inspired by the construction of \emph{Quasi-topological gravity} black holes in \cite{Quasi}. This consists in considering the action as a functional of these functions, $S[N,f]$. In fact, using the chain rule is it possible to show that the following equations hold for any higher-derivative Lagrangian 

\begin{equation}
\frac{1}{4\pi r^2}\frac{\delta S[N,f]}{\delta N}= \frac{ 2 \E_{tt}}{f N^2}\, , \quad
\frac{1}{4\pi r^2} \frac{\delta S[N,f]}{\delta f}= \frac{\E_{tt}}{N f^2 }+N \E_{rr}\, .
\end{equation}
 Here, $\mathcal{E}_{tt}$ and $\mathcal{E}_{rr}$ are the $tt$ and $rr$ components of the corresponding field equations\footnote{Note that $\mathcal{E}_{a b}=\frac{1}{\sqrt{|g|}}\frac{\delta S}{\delta g^{ab}}
$}. Therefore, one finds 
\begin{equation}
\frac{\delta S[N,f]}{\delta N}= \frac{\delta S[N,f]}{\delta f}=0 \Leftrightarrow \E_{tt}=\E_{rr}=0\, ,
\end{equation}
\ie imposing the variations of $S[N,f]$ to vanish is equivalent to imposing those components of the field equations to be solved. Finally, the Bianchi identity $\nabla^{a}\E_{ab}=0$ ensures that the angular equations also hold whenever $ \E_{tt}=\E_{rr}=0$ are satisfied. This shows that the equations for $f(r)$ and $N(r)$ can be obtained from the action functional $S[N,f]$  without need to compute the full non-linear equations explicitly.

Interestingly, for four-dimensional ECG \req{ECGaction}, the action functional $S[N,f]$ can be written as 
\begin{equation}
\label{Nfaction}
\begin{aligned}
S[N,f]=\frac{1}{8\pi G}\int dr &N(r) \cdot \bigg\{-\frac{1}{3}\Lambda_0r^3-(f-1)r-G^2 \lambda \bigg[4f'^3+12\frac{f'^2}{r}-24f(f-1)\frac{f'}{r^2}-12ff''\left(f'-\frac{2(f-1)}{r}\right)\bigg]\bigg\}'\\
&+\ldots\, ,
\end{aligned}
\end{equation}
where $()^{\prime}=d()/dr$ and where the ellipsis denote terms involving at least two derivatives of $N$, like $N'^2/N, N''N'/N$, $N'^3/N^2$, and so on. Now we can get the equations of $N$ and $f$ by computing the variation of this action with respect to them. Since $N$ is multiplied by a total derivative in (\ref{Nfaction}), when we compute $\delta_f S$ we get an expression which is homogeneous in derivatives of $N$. Hence, $\delta_f S=0$ can be solved by imposing 
\begin{equation}
N'(r)=0\, .
\end{equation}
This is enough to show that ECG admits solutions characterized by a single function $f(r)$. From now on we set $N=1$. On the other hand, the equation $\delta_N S=0$ yields, after setting $N=1$ and integrating once, the following equation for $f$:
\begin{equation}
\label{fequation}
\begin{aligned}
-\frac{1}{3}\Lambda_0r^3-(f-1)r-G^2 \lambda \bigg[4f'^3
+12\frac{f'^2}{r}-24f(f-1)\frac{f'}{r^2}
-12ff''\left(f'-\frac{2(f-1)}{r}\right)\bigg]=r_0\, ,
\end{aligned}
\end{equation}
where $r_0$ is an integration constant. Let us stress here two important points. In general, in higher-order gravities one cannot set $N$ to a constant, so the solutions are characterized by two different functions, see \eg \cite{Lu:2015cqa,Lu:2015psa}. 
Moreover, the equations of motion of higher-order gravity include in general up to fourth-order derivatives. Here we have been able to reduce the problem to a second-order equation for a single function. 

\section{Asymptotically flat black hole}\label{asy}
In this section we construct black-hole solutions of \req{ECGaction} using \req{fequation}. For simplicity, we focus on the asymptotically flat case, \ie we set $\Lambda_0=0$. Unfortunately, we have not been able to solve \req{fequation} analytically. However, we can make several expansions and approximations  which will help us understand the nature of the solution and will enable us to show that the first-law is exactly satisfied for our solutions. At the end of the section we also plot some numerical solutions of \req{fequation} corresponding to generalized Schwarzschild black holes for various values of the coupling $\lambda$.
\subsection{Asymptotic behavior}
Since (\ref{fequation}) is a second-order differential equation, it possesses a two-parameter family of solutions. 
We will require the solution to be asymptotically flat, so that 
\begin{equation}\label{bdy}
\lim_{r\rightarrow+\infty} f( r )=1\, .
\end{equation}
Then, the question is:  does this condition completely fix the solution? In order to answer it, we can make an expansion around $r\rightarrow +\infty$. We assume that, for $r\rightarrow +\infty$ the solution can be expressed as Schwarzschild plus a small correction, \ie
\begin{equation}
f( r )=1-\frac{r_0}{r}+f_1( r )\, ,
\end{equation}
where we assume that $| f_1( r )|\ll1$. Inserting this into (\ref{fequation}) and expanding linearly in $f_1$ we obtain a differential equation for the correction:
\begin{equation}
\begin{aligned}
-r^6 f_1-G^2\lambda (108 r_0^2-92 r_0^3/r)+12 G^2\lambda  r_0\Big[(6r-14 r_0)f_1
+3r(r_0-2r)f_1'+3r^2(r-r_0)f_1''\Big]=0\,.
\end{aligned}
\end{equation}
The general solution of the above equation is given by the sum of the homogeneous solution plus a particular solution, $f_1=f_{1,p}+f_{1,h}$. To first order in $\lambda$, a particular solution is
\begin{equation}\label{party}
f_{1,p}( r )=G^2 \lambda\left(-\frac{108 r_0^2}{r^6}+\frac{92 r_0^3}{r^7}\right)+\mathcal{O}\left(\lambda^2,\frac{r_0^4}{r^8}\right)\, ,
\end{equation}
where terms with higher orders in $\lambda$ decay faster as $r\rightarrow+\infty$, so the first term provides a good approximation. The homogeneous equation can in turn be written as:
\begin{equation}
f_{1,h}''-\gamma ( r ) f_{1,h}' -\omega^2( r )f_{1,h}=0\, ,\quad \text{where} \quad \omega^2( r )=\frac{r^4}{36G^2\lambda r_0(r-r_0)}-\frac{6r-14 r_0}{3r^2(r-r_0)}, \quad \gamma( r )=\frac{2r-r_0}{r(r-r_0)}\, .
\end{equation}
Now, when $r$ is large we get $\omega'/\omega^2\ll1$ and $\gamma\ll\omega$. In this situation, the solution of the previous equation is approximately $f_{1,h}\approx A \exp\left[\int dr \omega( r )\right]+B\exp\left[-\int dr \omega( r )\right]$, for arbitrary constants $A$ and $B$. In particular, when  $r\rightarrow+\infty$, we get $\omega^2=r^3/(36G^2\lambda r_0)+\mathcal{O}(r^2)$, and the solution is given very approximately by
\begin{equation}
f_{1,h}( r )\simeq A \exp\left(\frac{r^{5/2}}{15G\sqrt{\lambda r_0}}\right)+B \exp\left(-\frac{r^{5/2}}{15G\sqrt{\lambda r_0}}\right)\, .
\label{homogeneous}
\end{equation}
Now, since we want the metric to be asymptotically flat, we must set $A=0$. This leaves us with a 1-parameter family of solutions which are asymptotically flat. Hence, in the large $r$ limit, the solution is given by
\begin{equation}
f( r )\simeq 1-\frac{r_0}{r}-G^2\lambda\left(\frac{108 r_0^2}{r^6}-\frac{92 r_0^3}{r^7}\right)+\mathcal{O}\left(\lambda^2,\frac{r_0^4}{r^8}\right)+B  \exp\left(-\frac{r^{5/2}}{15G \sqrt{\lambda r_0}}\right)\, ,
\label{asymptotic}
\end{equation}
for some constant $B$. Observe that all the leading asymptotic corrections to the Schwarzschild metric come from the solution \req{party}, while the contributions from the homogeneous equation are extremely subleading. Hence, the term proportional to $B$ above can be discarded from the asymptotic expansion \req{asymptotic}.
Note that had we considered a theory with massive modes (of mass $m$) in the linearized spectrum, we would have expected the corresponding asymptotic expansion to contain decaying exponential terms $\sim e^{-m r}$. By construction, ECG does not propagate massive modes linearly on the vacuum \cite{PabloPablo} and, consistently, those terms do not appear in \req{asymptotic}. However, since ECG is a higher-derivative theory, nothing prevents additional pseudo-modes from appearing at the non-linear level, or on backgrounds different from the vacuum. This seems to be the case here, because we can associate the decaying exponential in (\ref{asymptotic}) with a pseudo-mode of mass $m^2=\omega^2( r )$. Indeed, the decay is faster-than-exponential because the mass of this pseudo-mode goes to infinity as $r\rightarrow\infty$. Hence we see that, even though ECG can propagate additional pseudo-modes in backgrounds different from the vacuum, these modes are rapidly killed in the asymptotic limit, so they only live in a bounded region. The same behavior is expected to occur for any other higher-order gravity which only propagates a massless graviton on the vacuum --- \ie those belonging to the \emph{Einstein-like} class in the classification of \cite{Bueno:2016ypa}. 
From the asymptotic expansion above we can obtain the mass of the black hole.
  In the case of an asymptotically flat space-time, the Abbott-Deser mass formula is not changed by higher-order curvature terms, so we can apply the usual recipe \cite{Abbott:1981ff,Deser:2002jk}. In particular, the total mass can be found in our case through
  \begin{equation}
  M=\frac{1}{2G}\lim_{r\rightarrow+\infty}r(g_{rr}( r )-1)\, .
 \end{equation} 
  Now, as we said before, higher-order corrections in $\lambda$ will decay with higher powers of $r$ as $r\rightarrow +\infty$, so the leading term, $-r_0/r$, will not be affected by these corrections. Therefore, this formula yields
\begin{equation}\label{massro}
r_0=2GM\, ,
\end{equation}
as usual. Naturally, using this and $\req{asymptotic}$ we can write the final expression for the asymptotic expansion of $f(r)$ --- for small values of $\lambda$ --- as
\begin{equation}
f( r\rightarrow \infty )= 1-\frac{2GM}{r}-G^2\lambda\left(\frac{108 (2GM)^2}{r^6}-\frac{92 (2GM)^3}{r^7}\right)\, .
\label{asymptotic2}
\end{equation}

\subsection{Horizon}\label{hori}
For a metric of the form (\ref{ansatz0}), a horizon is a surface $r=r_h$ at which $f(r_h)=0$ and $f'(r_h)\ge 0$. In particular, the function must be differentiable at $r_h$. Note also that the surface gravity on the horizon for this kind of metric is just $\kappa_g=f'(r_h)/2$. Assuming that the function $f$ is completely regular at the horizon and that it can be Taylor-expanded around it \footnote{As we see from (\ref{fequation}), at the horizon --- \ie when $f=0$ --- the term which multiplies $f''$ vanishes. This can give rise to non-differentiability on the horizon of some of the solutions, so imposing that the horizon is regular is indeed a strong restriction.}, we can write
\begin{equation}\label{Hexpansion}
f( r )=2 \kappa_g (r-r_h)+\sum_{n=2}^{\infty} a_n(r-r_h)^n\, ,
\end{equation}
where we have made explicit the first term, and where $a_n=f^{(n)}(r_h)/n!$.
The idea is to plug this expansion into (\ref{fequation}) and solve order by order in $(r-r_h)^n$. Up to quadratic order we get
\begin{align}
&+r_h-2GM-16\lambda\kappa_g^2\left(2\kappa_g+\frac{3}{r_h}\right)
+\left(1-2\kappa_g r_h-48\lambda\frac{\kappa_g^2}{r_h^2}\right)(r-r_h)\\ \notag &+
\Bigg[\lambda  \Big(144 a_3 \kappa _g\left(\kappa_g+\frac{1}{r_h}\right)+48 a_2^2 \kappa _g-\frac{144 a_2 \kappa _g}{r_h^2}
-\frac{192 a_2 \kappa _g^2}{r_h}+\frac{144 \kappa _g^2}{r_h^3}+\frac{192 \kappa _g^3}{r_h^2}\Big)
-a_2r_h-2 \kappa _g\Bigg](r-r_h)^2\\ \notag &+\mathcal{O}((r-r_h)^3)=0,
\end{align}
where we have already taken into account \req{massro}. Now this equation must hold at every order in $(r-r_h)$. We see that the first two equations determine the horizon radius $r_h$ and the surface gravity $\kappa_g$ as a function of the mass,
\begin{eqnarray}
\label{rh}
r_h-2GM-16\lambda\kappa_g^2\left(2\kappa_g+\frac{3}{r_h}\right)&=&0\, ,\\
\label{kg1}
1-2\kappa_g r_h-48\lambda\frac{\kappa_g^2}{r_h^2}&=&0\, .
\end{eqnarray}
It is important to stress that the above expressions are exactly true. The fact that we can obtain exact relations among the mass, the horizon radius and the surface gravity is remarkable and not shared by other higher-order gravities.
Once $r_h$ and $\kappa_g$ are determined, from the third equation we get a relation between $a_2$ and $a_3$. Since it is linear in $a_3$, we can easily determine $a_3$ as a function of $a_2$. In the fourth equation $a_4$ appears linearly, so we can obtain it as a function of the previous coefficients, and so on. In general, from the $n-$th equation we can determine the coefficient $a_n$. Hence, we get a family of solutions with only one free parameter, $a_2$. We learn two things from this: one is that the theory \req{ECGaction} admits black hole solutions with regular horizons. Another is that the regular horizon condition reduces the number of solutions from a two-parameter family to a one-parameter one.

The expressions (\ref{rh}) and (\ref{kg1}) allow us to determine $\kappa_g$ and $r_h$ as  functions of the mass. However, it is easier to obtain the relations $\kappa_g(r_h)$ and $M(r_h)$. We get \footnote{In fact there is another solution with a minus sign in front of the square root, but that choice does not reproduce the correct limit when $\lambda=0$. }
\begin{equation}
\kappa_g=\frac{1}{r_h(1+\sqrt{1+48G^2\lambda/r_h^4})}\, , \quad 
\label{surfacegravity}
\frac{2GM}{r_h}=1-\frac{16G^2\lambda}{r_h^4}\frac{\Big(5+3\sqrt{1+48G^2\lambda/r_h^4}\Big)}{\Big(1+\sqrt{1+48G^2\lambda/r_h^4}\Big)^{3}}  \, .
\end{equation}
In Fig. \ref{fig2} we plot $M(r_h)$. From this it is obvious that all solutions have a greater horizon radius than Schwarzschild, \ie
\begin{equation}
r_{h}(\lambda)\geq r_h(0) \, \quad \text{for all} \quad \lambda \geq 0\, .
\end{equation}

 \begin{figure}[ht!]
\centering 
\includegraphics[scale=0.56]{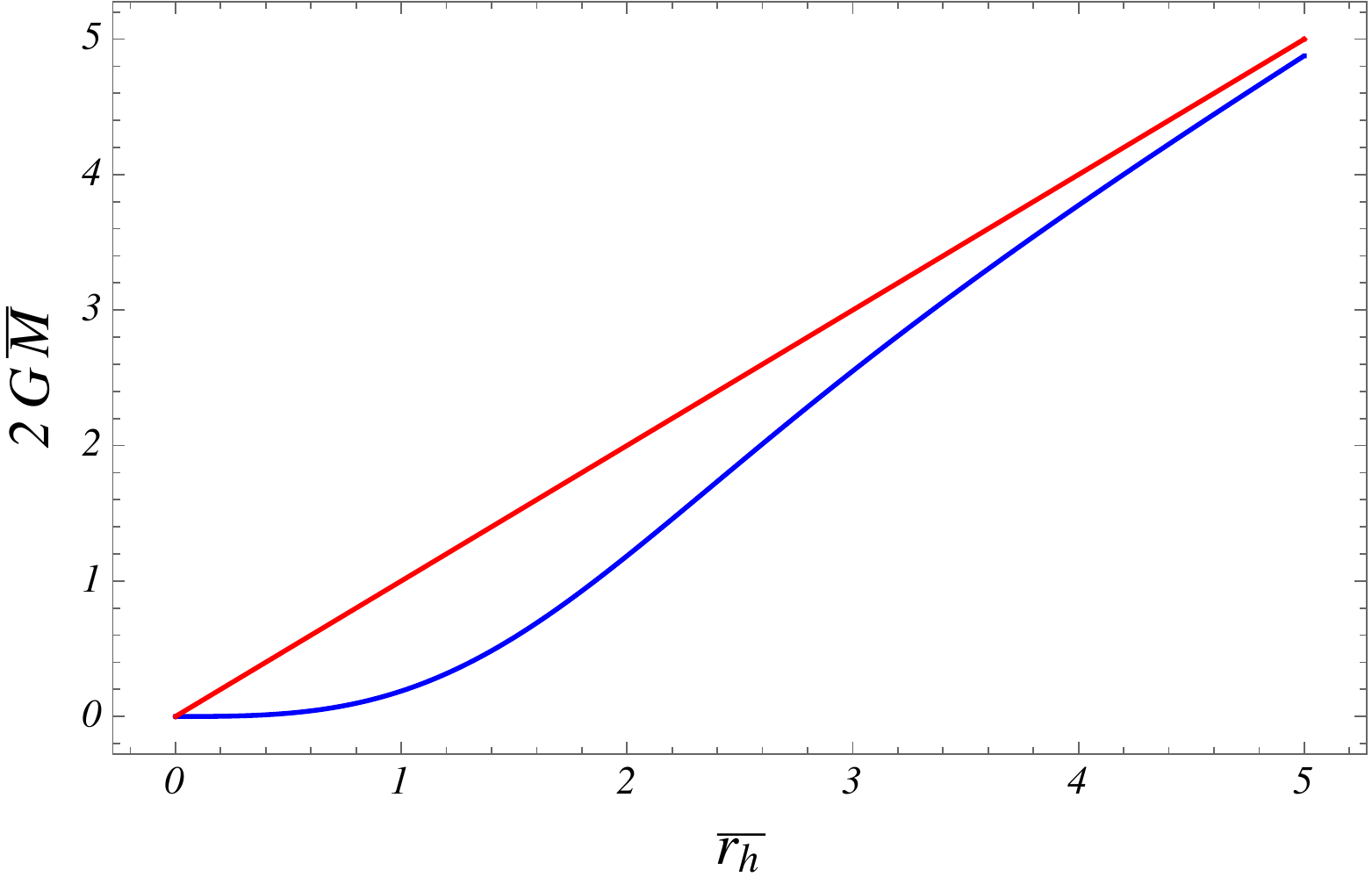}
\caption{We plot $2G\bar{M}$ as a function of $\bar{r}_h$ for the ECG solution (blue) with $\lambda>0$ and the usual Schwarzschild solution (red) where, for the sake of clarity, we defined $\bar{M}=M/(G^2 \lambda)^{1/4}$ and $\bar{r}_h=r_h/(G^2 \lambda)^{1/4}$. Note that the blue plot is valid for all values of $\lambda> 0$.} 
\labell{fig2}
\end{figure}

\subsection{Numerical solution}
In the previous two subsections we have argued that we can construct a one-parameter family of asymptotically flat solutions and another one-parameter family of solutions which posseses a regular horizon. Thus, we expect that there exists one solution which connects both.  A numerical computation shows that this is in fact the case. In order to perform the numerical computation, we start from the solution at the horizon, with $f(r_h)=0$, $f'(r_h)=2 \kappa_g$, with both $r_h$ and $\kappa_g$ determined in terms of the mass and $\lambda$ through (\ref{rh}) and (\ref{kg1}), and then we choose the following value for the free parameter, \begin{equation}
a_2=f''(r_h)/2\, .
\end{equation}
 This must in fact be chosen very carefully so that we do not excite the growing exponential mode in (\ref{homogeneous}). For that value of $a_2$, we are able to construct numerically the solution up to a sufficiently large $r$ for which the solution becomes very similar to Schwarzschild. Then, for larger $r$  the approximation (\ref{asymptotic}) holds and we can use it to continue the solution all the way to $r=+\infty$. Also, since the horizon is regular, we can in practice continue the solution to the inner region $f<0$. The result for various values of $G^2 \lambda$ is presented in Fig. \ref{fig1}.
 \begin{figure}[ht!]
\centering 
\includegraphics[scale=0.71]{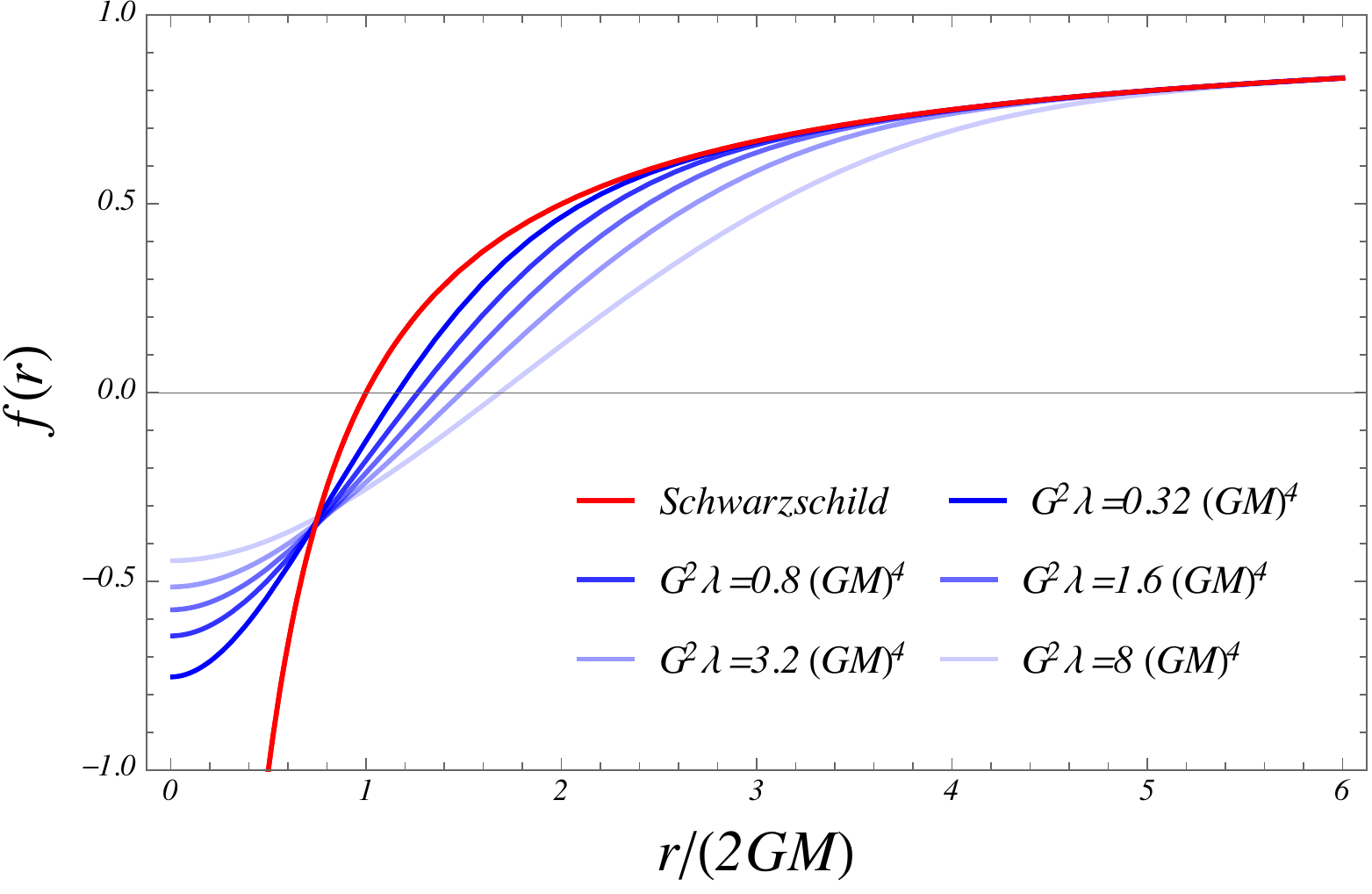}
\caption{Profile of $f( r )$ for several values of $\lambda$. The red line corresponds to the usual Schwarzschild blackening factor, $\lambda=0$. } 
\labell{fig1}
\end{figure}

As we can see, the solutions are very similar to Schwarzschild when $r$ is large enough, but they differ notably as $r$ approaches the horizon. Note again that the horizon radius of all solutions with $\lambda>0$ is greater than the Schwarzschild value $r_h=2GM$. Remarkably, $f(r)$ does not diverge at the origin $r=0$. In fact, it can be shown that the behaviour at $r=0$ is approximately $f( r )=a+b r^2+\mathcal{O}(r^3)$, for some constants $a$ and $b$.  Although there is no metric divergence, the curvature is still divergent at the origin. The divergence gets softened with respect to the Schwarzschild case though. In particular, the Kretschmann scalar reads 
\begin{equation}\label{krets}
R_{abcd}R^{abcd}=\frac{4(f(0)-1)^2}{r^4}+\mathcal{O}\left(\frac{1}{r^2}\right)\, ,
\end{equation}
where $f(0)$ is the constant value which $f(r)$ takes at $r=0$ when $\lambda>0$. Note that the limit $\lambda \rightarrow 0$ is not continuous in the above expression because $f(0)$ diverges for the Schwarzschild solution. In that case, one finds the usual result $R_{abcd}R^{abcd}=48G^2M^2/r^6$ instead. Note also that if we had $f(0)=1$, the singularity would be completely removed --- the $\mathcal{O}(1/r^2)$ term would also vanish. Although this never happens for the ECG black holes, since we always have $f(0)<0$, it could be possible that the addition of even higher-order terms, --- \eg quartic ones --- would completely remove the singularity.


\section{Black Hole Thermodynamics}\label{them}
Our analysis in section \ref{hori} allowed us to obtain the horizon properties $r_h$ and $\kappa_g$ as functions of the black-hole mass $M$ and $\lambda$. As we stressed, those results are exact, \ie fully non-perturbative in $\lambda$. Let us now study some thermodynamic properties \cite{Hawking:1974sw,Bardeen:1973gs,Bekenstein:1973ur,Bekenstein:1974ax} associated to these solutions. According to Wald's formula \cite{Wald:1993nt,Iyer:1994ys,Jacobson:1993vj}, the entropy of a black hole in a higher-order derivative theory of gravity is given by
\begin{equation}
\mathsf{S}=-2\pi \int_{H} d^2x\sqrt{h} \frac{\delta \mathcal{L}}{\delta R_{abcd}}\epsilon_{ab}\epsilon_{cd}\, ,
\end{equation}
where $\frac{\delta }{\delta R_{abcd}}$ is the Euler-Lagrange derivative, $\mathcal{L}$ is the gravitational Lagrangian, $h$ is the determinant of the induced metric on the horizon and $\epsilon_{ab}$ is the binormal of the horizon, normalized as $\epsilon_{ab}\epsilon^{ab}=-2$. 

Let us now apply this formula to our theory \req{ECGaction}. At this point it is convenient for our purposes to turn on an explicit Gauss-Bonnet term in the action, \ie $\mathcal{L} \rightarrow \mathcal{L}+ \frac{\alpha}{16 \pi} \mathcal{X}_4$, where  $\mathcal{X}_4=R^2-4R_{ab}R^{ab}+R_{abcd}R^{abcd}$. This term has no effect in our discussion so far, but we will make use of it below. The result for $\mathsf{S}$ reads
\begin{equation}
\begin{aligned}
\mathsf{S}=\frac{1}{4G} \int_{H} d^2x\sqrt{h}\Big[&1+2\alpha G R_{(2)}+G^2\lambda\Big(36 R_{b\ d}^{\ e\ f}R_{aecf}+3R_{ab}^{\ \ ef}R_{cdef}\\
&-12 R_{ac}R_{db}-24R^{ef}R_{ebfc}g_{bd}+24g_{bd}R_{ce}R^{e}_{\ a}\Big)\epsilon^{ab}\epsilon^{cd}\Big]\, ,
\end{aligned}
\end{equation}
where $R_{(2)}$ is the Ricci scalar of the induced metric in the horizon, coming from the Gauss-Bonnet term. Naturally, this term yields a topological contribution.
For a metric of the form (\ref{ansatz0}) and with a spheric horizon placed at $r=r_h$, one finds
\begin{equation}
\label{GeneralEntropy}
\mathsf{S}=\frac{\pi r_h^2}{G}\left[1-48G^2 \lambda \frac{\kappa_g^2}{r_h^2}\left(\frac{2}{\kappa_g r_h}+1\right)\right]+2\pi\alpha\, ,
\end{equation}
where we have taken into account that $f'(r_h)=2\kappa_g$. This expression for the entropy is in principle valid for any static and spherically symmetric black hole solving the equations of motion of \req{ECGaction}. For the black hole we have constructed, the surface gravity is given by (\ref{surfacegravity}), so we can write the entropy in terms of the radius as
\begin{equation}
\mathsf{S}=\frac{\pi r_h^2}{G}\left[1- \frac{48G^2\lambda}{r_h^4}\frac{\left(3+2\sqrt{1+48G^2\lambda/r_h^4}\right)}{\left(1+\sqrt{1+48G^2\lambda/r_h^4}\right)^{2}}\right] +2\pi\alpha\, .
\label{Entropy}
\end{equation}
Also, by using the relation between $M$ and $r_h$ in (\ref{surfacegravity}), we can obtain the relation $\mathsf{S}(M)$ parametrized by $r_h$.  
For $M=0$ the entropy reads $\mathsf{S}(M=0)=-2\pi\sqrt{48\lambda}+2\pi\alpha$. Hence, we can fix the Gauss-Bonnet coupling to $\alpha=\sqrt{48\lambda}$ so that the horizon entropy vanishes when $M=0$. As is clear from Fig. \ref{fig4}, the entropy is positive and larger than the Schwarzschild one for all other values of $M$ and $\lambda\geq 0$. 
\begin{figure}[h]
        \centering
   		 \subfigure[ ]{
                \includegraphics[scale=0.52]{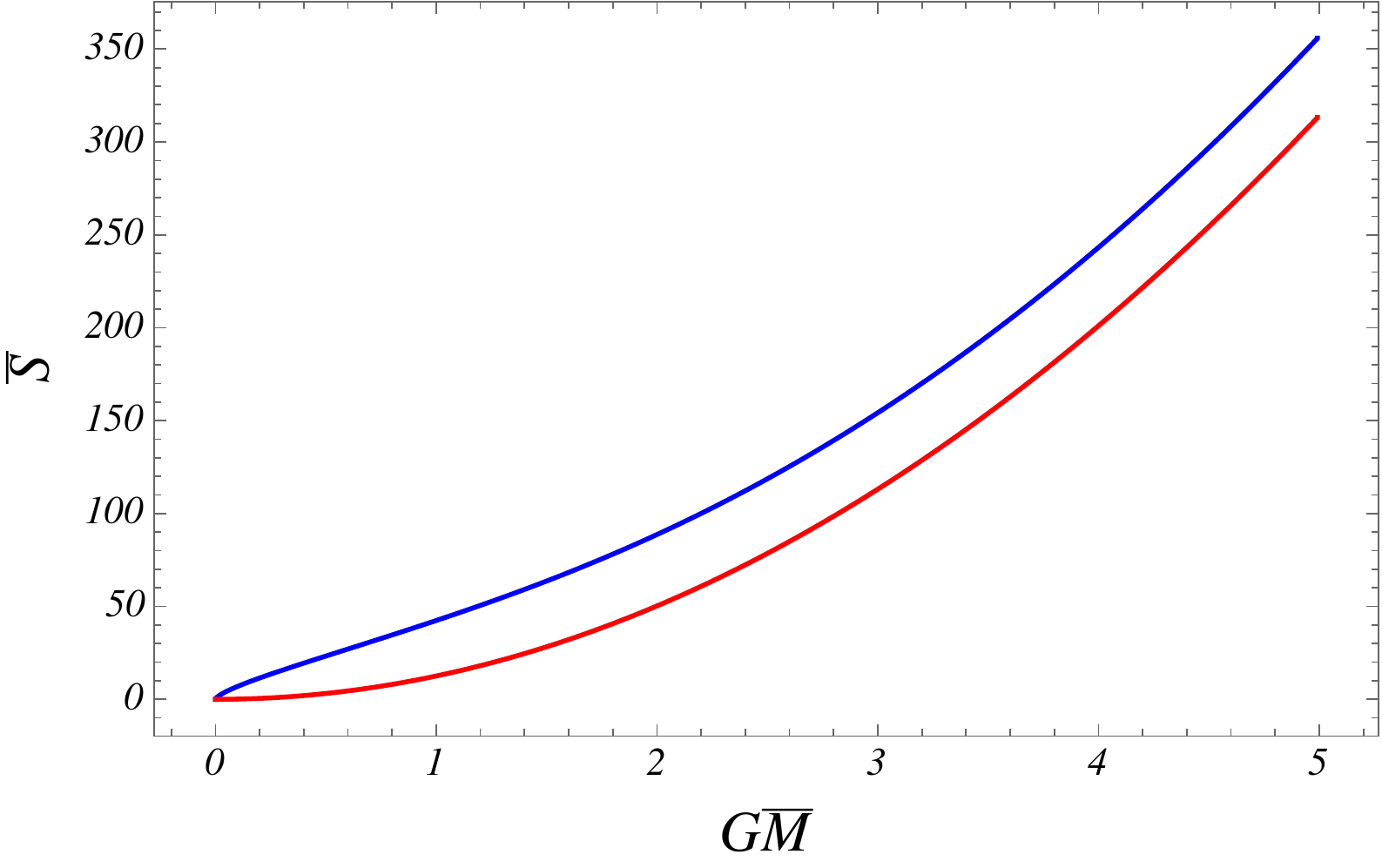}}
                \ \ \ 
         \subfigure[ ]{
                \includegraphics[scale=0.54]{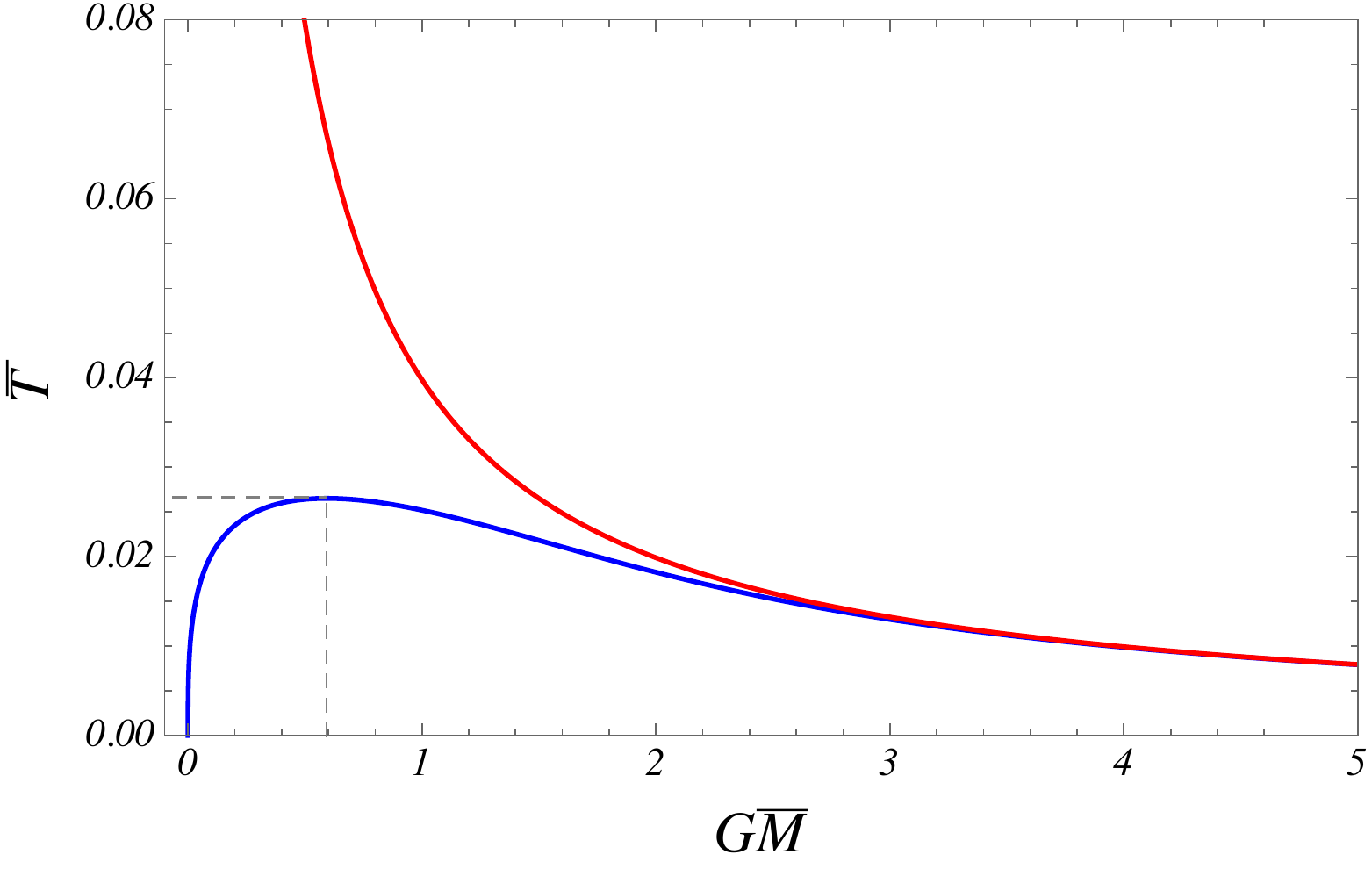}}
        \caption{(a) We plot $\mathsf{S}(M)$ for the ECG black hole with $\lambda>0$ (blue) and the usual Schwarzschild solution (red). Again, we use the normalized mass $\bar{M}=M/(G^2 \lambda)^{1/4}$ and also $\bar{S}=S/ \lambda^{1/2}$. b) We plot the Hawking temperature as a function of its mass for Schwarzschild (red) and the ECG solution (blue). In this case we used the normalized temperature $\bar{T}=T\cdot (G^2\lambda)^{1/4}$. The dashed lines highlight the presence of a maximum temperature --- see \req{tee}. For smaller values, there exist two solutions for each $T$.}
\labell{fig4}
\end{figure}

On the other hand, the Hawking temperature \cite{Hawking:1974sw} of our solution can be written in terms of the radius as 
\begin{equation}
T=\frac{1}{2\pi r_h(1+\sqrt{1+48G^2\lambda/r_h^4})}\, ,
\label{Temperature}
\end{equation}
where we used  \req{surfacegravity}. 
This temperature increases as the mass decreases up to a maximum temperature 
\begin{equation}\label{tee}
T_{\rm{max}}=\frac{1}{12\pi G^{1/2}  \lambda^{1/4}}\, , \quad \text{which is reached for a mass}\quad M_{\rm{max}}=16/27 \cdot G^{-1/2}\lambda^{1/4}\, .
\end{equation}
Then, the temperature decreases until it vanishes for $M=0$. As we can see from Fig. \ref{fig4}, this behavior is very different from the Einstein gravity one, since for this the temperature blows up as $M\rightarrow 0$. In fact, the temperature behaves in a similar fashion to the one of the usual Reissner-Nordstr\"om (RN) solution to the Einstein-Maxwell system --- see \eg \cite{Ortin:2004ms}. In that case, the temperature also reaches a maximum value $T\sim 1/|Q|$ and vanishes as $M\rightarrow +\infty$. An important difference with respect to that case is that for RN, the temperature vanishes when the extremality condition $M^2=Q^2$ is met, \ie for a positive value of the mass, whereas for the ECG black holes this occurs when the mass goes to zero.

Now, using (\ref{Temperature}) and (\ref{Entropy}) it is possible to show that the First law of black hole mechanics
\begin{equation}
dM=T d\mathsf{S}\, ,
\end{equation}
holds exactly.
This is an interesting check of our calculations, since the three physical quantities appearing in this expression --- namely, the Abbott-Deser mass $M$, the Wald entropy $\mathsf{S}$ and the Hawking temperature $T$ --- have been computed independently.

Observe also that using \req{tee} and \req{Entropy} it is possible to find the following explicit expression for the entropy as a function of the temperature
\begin{equation}
\mathsf{S}/\mathsf{S}_{E}=\frac{(T/T_{E})^2-4(1-T/T_{E})+\sqrt{1-T/T_{E}}}{T/T_{E}}\, ,
\end{equation}
where $\mathsf{S}_{E}=\pi r_h^2/G$ and $T_{E}=1/(4\pi r_h)$ are the Einstein gravity values of the entropy and the temperature, \ie those corresponding to the Schwarzschild solution. When $T=T_{E}$, one recovers $\mathsf{S}=\mathsf{S}_{E}$, as expected.

 We can also compute the specific heat, defined as
\begin{equation}
C=T\left(\frac{\partial \mathsf{S}}{\partial T}\right)_M.
\end{equation}
Parametrized in terms of $r_h$, we can write it as
\begin{equation}\label{cct}
C=-\frac{8 \pi  \left(1152 G^4\lambda ^2+24 G^2\lambda  r_h^4 \left(4 \sqrt{1+\frac{48 G^2\lambda }{r_h^4}}+5\right)+r_h^8 \left(\sqrt{1+\frac{48 G^2\lambda }{r_h^4}}+1\right)\right)}{Gr_h^2 \left(\sqrt{1+\frac{48 G^2\lambda }{r_h^4}}+1\right)^2 \left(-48G^2 \lambda +r_h^4 \left(\sqrt{1+\frac{48G^2 \lambda }{r_h^4}}+1\right)\right)}\, .
\end{equation}
It has two regions: for $M>M_{\rm{max}}$, it is negative as in the Schwarzschild black hole, while for $M<M_{\rm{max}}$ we get $C>0$. It diverges in the midpoint $M=M_{\rm{max}}$. As we can see from Fig. \req{fig6}, when expressed as a function of the temperature, $C(T)$ has two branches, one negative and one positive, and both diverge at $T=T_{\rm{max}}$, which suggests the presence of a phase transition.
 \begin{figure}[ht!]
\centering 
\includegraphics[scale=0.66]{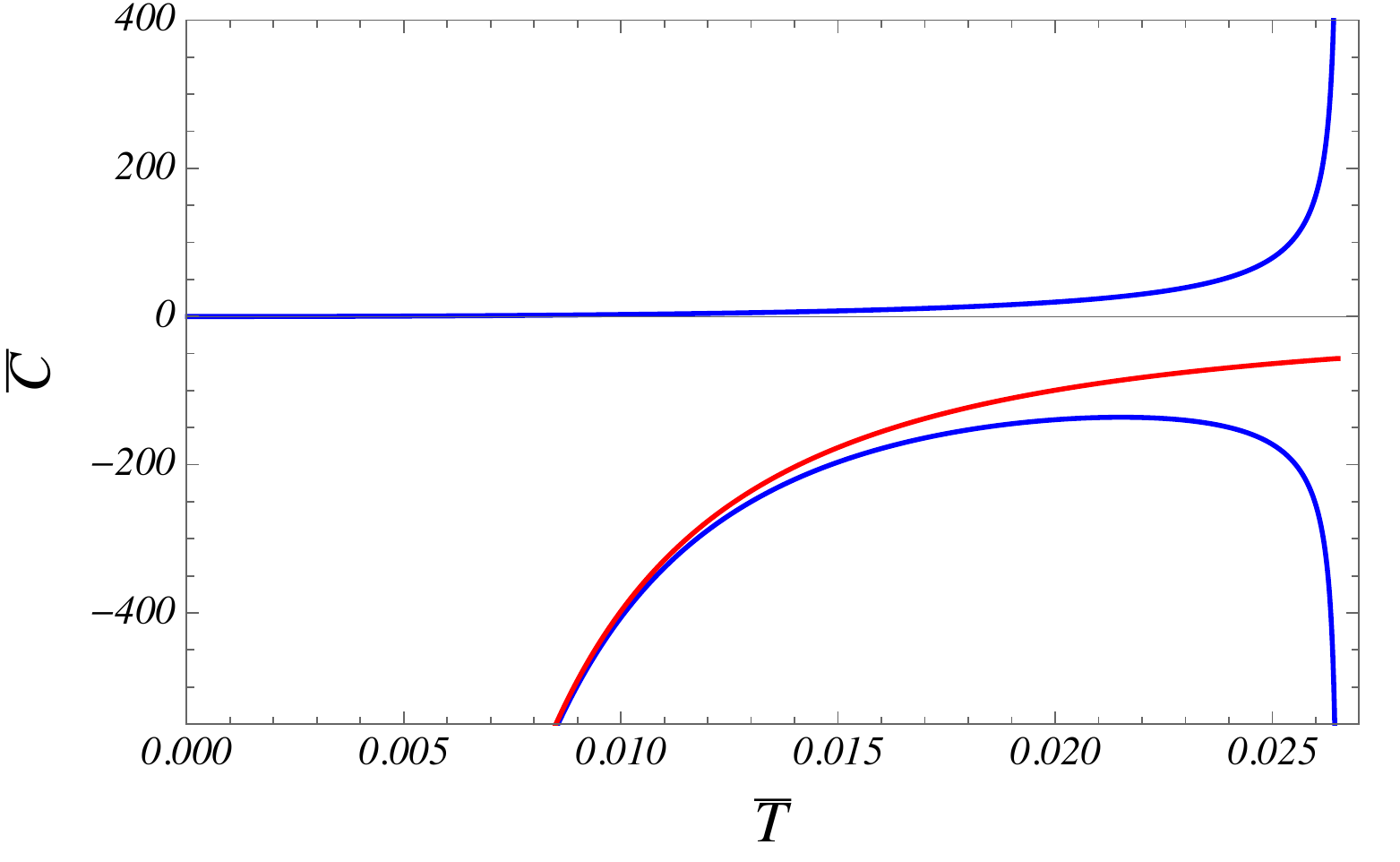}
\caption{We plot $C$ as a function of the temperature for the ECG solution (blue) with $\lambda>0$ and the usual Schwarzschild solution (red). In this case, we defined the normalized specific heat and temperature as $\bar C=C/ \lambda^{1/2}$ and again $\bar{T}=T\cdot (G^2\lambda)^{1/4}$.} 
\labell{fig6}
\end{figure}
In Fig. \req{fig6}, the upper blue branch corresponds to the solutions with $M<M_{\rm{max}}$, while the ones with $M>M_{\rm{max}}$ are the ones in the lower branch. The solutions with positive heat are thermodynamically stable, and very different from the usual Schwarzschild solution, which has $C(T)<0$ for all $T$ as is clear from Fig. \ref{fig6}.  As we saw, for a given temperature $T$ there exist two solutions with different horizon radius --- and hence mass, entropy, etc. In all cases, the one with the smaller $r_h$ is the one with positive specific heat, and vice-versa. The situation is reminiscent to the one observed in \cite{Myers:1988ze}, where certain odd-dimensional Lovelock black holes were shown to become stable for small enough masses.

\section{Generalized Reissner-Nordstr\"om-(Anti-)de Sitter  solution}\label{charged}
In the previous section we focused on the simplest possible case, corresponding to generalized versions of the asymptotically flat Schwarzschild black hole. However, these solutions can be easily generalized. Here we will turn on the cosmological constant $\Lambda_0$ and add a Maxwell field to the action \req{ECGaction}. These two extensions allow us to obtain generalized versions of the usual Reissner-Nordstr\"om-(Anti-)de Sitter (RN-(A)dS) black hole. Hence, let us consider now the action
\begin{equation}
S=\int_{\mathcal{M}}d^4x\sqrt{|g|}\left[\frac{1}{16 \pi G}\left(-2\Lambda_0+R+\alpha G \mathcal{X}_4-\lambda G^2\mathcal{P}\right)-\frac{1}{4}F_{ab}F^{ab}\right],
\end{equation}
where $F_{ab}=2\partial_{[a}A_{b]}$. We consider the same ansatz \req{ansatz1} for the metric, while for the vector field we choose
\begin{equation}
A=A_0(r) dt	\,.
\end{equation}
As before, we find that $N'=0$, so we set $N(r)=1$. On the  other hand, we find the following equation for $A_0$,
\begin{equation}
-\left(A_0'\frac{r^2}{N(r)}\right)'=0\, .
\end{equation}
Since $N(r)=1$, this equation yields the usual expressions 
\begin{equation}\label{elec}
A_0=\frac{q}{4 \pi r}\, , \quad F=\frac{q}{4\pi r^2}dt\wedge dr\, ,
\end{equation}
for the electric potential and its field strength. Here, $q$ is an integration constant related to the electric charge of the solution. Finally, the equation for $f$ is found from the variation with respect to $N$. It reads
\begin{equation}\label{fist}
-(f-1)r-G^2 \lambda \bigg[4f'^3
+12\frac{f'^2}{r}-24f(f-1)\frac{f'}{r^2}
-12ff''\left(f'-\frac{2(f-1)}{r}\right)\bigg]=\frac{1}{3}\Lambda_0 r^3+r_0-\frac{G Q^2}{r},
\end{equation}
where $Q^2=q^2 /(4\pi)$, and $r_0$ is an integration constant. Observe that for $\lambda=0$ we obtain
\begin{equation}
f(r)=1-\frac{\Lambda_0r^2}{3}-\frac{2GM}{r}+\frac{GQ^2}{r^2},
\end{equation}
where we identified $r_0=2GM$.  This is of course nothing but the usual RN-(A)dS blackening factor. Interestingly, when $\lambda$ is turned on, the asymptotic quantities get corrected in this case. Indeed, by performing an asymptotic expansion around $r\rightarrow +\infty$, we see that $r_0$ is identified with the mass as before, $r_0=2 GM$, and that $f(r)$ takes the form 
\begin{equation}
f(r)=1-\frac{1}{3}\Lambda_{\rm{eff}}r^2-\frac{2 G_{\rm{eff}}M}{r}+\frac{G_{\rm{eff}} Q^2}{r^2}+\mathcal{O}\left(\frac{1}{r^3}\right),
\end{equation}
where the effective cosmological constant $\Lambda_{\rm{eff}}$ is a solution of the equation
\begin{equation}\label{geg}
\frac{16}{9} \lambda G^2 \Lambda_{\rm{eff}}^3-\Lambda_{\rm{eff}}+\Lambda_0=0\,,
\end{equation}
and where the effective gravitational constant is given by
\begin{equation}
G_{\rm{eff}}=\frac{G}{1-\frac{16}{3}\lambda G^2 \Lambda_{\rm{eff}}^2}\,.
\end{equation}
Observe that \req{geg} is nothing but the embedding equation that pure (A)dS$_4$ with curvature $\Lambda=\Lambda_{\rm eff}/3$ must satisfy in order for it to be a solution of the ECG theory \cite{PabloPablo}.
 As in the uncharged asymptotically flat case, it is possible to compute higher-order terms in the asymptotic expansion. Similarly, one can study the thermodynamic properties of these black holes by making a Taylor expansion around the horizon, as in (\ref{Hexpansion}). In that case we find the following generalized equations, which relate $r_h$ and $\kappa_g$ to $M$, $Q$ and $\Lambda_0$:
 \begin{eqnarray}\label{ff}
-2GM+r_h-16\lambda G^2\kappa_g^2\left(2\kappa_g+\frac{3}{r_h}\right)-\Lambda_0\frac{r_h^3}{3}+\frac{GQ^2}{r_h}&=&0\, ,\\ \label{ff2}
1-\frac{GQ^2}{r_h^2}-\Lambda_0 r_h^2-2\kappa_g r_h-48\lambda G^2\frac{\kappa_g^2}{r_h^2}&=&0\, .
\end{eqnarray}
Using these relations, we can write the Hawking temperature $T=\kappa_g/(2\pi)$, the entropy $\mathsf{S}$ (\ref{GeneralEntropy}) and the mass $M$ in terms of $r_h$ and $Q$. The result reads
\begin{align}
T=&\frac{r_h^2-\Lambda_0 r_h^4-GQ^2}{2 \pi\left(r_h^3+  \sqrt{r_h^6+48G^2 \lambda ( r_h^2-\Lambda_0 r_h^4-GQ^2)}\right)}\, ,\\
\mathsf{S}=&\frac{\pi r_h^2}{G}\left[1-\frac{48 \lambda G^2 \left(r_h^2-\Lambda_0 r_h^4-GQ^2\right) \left(3 r_h^3-\Lambda_0 r_h^5-GQ^2 r_h+2 \sqrt{r_h^6+48G^2 \lambda ( r_h^2-\Lambda_0 r_h^4-GQ^2)}\right)}{r_h^3 \left(\sqrt{r_h^6+48G^2 \lambda ( r_h^2-\Lambda_0 r_h^4-GQ^2)}+r_h^3\right){}^2}\right]\\ \notag &+2\pi\alpha\, ,\\
\frac{2GM}{r_h}= &1+\frac{GQ^2}{r_h^2}-\frac{\Lambda_0 r_h^2}{3}\\ \notag &-\frac{16 G^2\lambda  \left(r_h^2-\Lambda_0 r_h^4-GQ^2\right)^2 \left(5 r_h^3-2 \Lambda_0 r_h^5-2 GQ^2 r_h+3 \sqrt{r_h^6+48G^2 \lambda ( r_h^2-\Lambda_0 r_h^4-GQ^2)}\right)}{r_h^2\left(\sqrt{r_h^6+48G^2 \lambda ( r_h^2-\Lambda_0 r_h^4-GQ^2)}+r_h^3\right)^3}\, . \hspace{0.7cm}
\end{align}
The last equation fixes the horizon radius $r_h$ in terms of $M$, $Q$ and $\Lambda_0$. Using these expressions,
it is possible to show --- using Mathematica --- that the first law also holds for these solutions. In this case, it reads
\begin{equation}
dM= T d\mathsf{S}+\Phi dq,
\end{equation}
where $\Phi=A_0$ is the electrostatic potential \req{elec}.  Let us also mention that the extremal limit --- corresponding to $\kappa_g=0$ --- coincides with the Einstein gravity one. This can be easily seen from  \req{ff} and \req{ff2}, which become in that case
\begin{equation}
1-\frac{2GM}{r_h}+\frac{GQ^2}{r_h^2}-\frac{\Lambda_0 r_h^2}{3}=0\, , \quad \Lambda_0 r_h^4-r_h^2+GQ^2=0\, ,
\end{equation}
the second of which imposes $T=0$, $\mathsf{S}=\pi r_h^2/G$.
This is because when $\kappa_g=0$, all terms involving $\lambda$ vanish in the previous equations. Therefore, neither the extremality condition nor the horizon radius are altered by the ECG term in that case.  In particular, if $\Lambda_0=0$, extremality is reached when $Q^2=GM^2$ and in that case, $r_h=\sqrt{G} |Q|$.


\section{Discussion}
In this paper we have constructed generalizations of the Schwarzschild and Reissner-Nordstr\"om black holes in four-dimensional Einsteinian cubic gravity \req{ECGaction} both with Minkowski and (A)dS asymptotes. We have shown that the theory admits solutions with a single function $f(r)$ determined through a non-linear second-order differential equation \req{fist} and studied some of their thermodynamic properties which, remarkably enough, can be accessed analytically. As far as we know, the new solutions represent the first non-trivial four-dimensional generalizations of the Schwarzschild- and RN-(A)dS black holes in higher-order gravity whose thermodynamic properties can be computed exactly.
Using those results we have been able to check analytically that the solutions satisfy the first law of black hole mechanics. 

We have observed that the addition of the ECG term to the EH action softens the black-hole singularity --- see \req{krets}. In particular, the metric of the ECG black hole does not diverge at $r=0$ in Schwarzschild coordinates and the Kretschmann  invariant diverges as $\sim r^{-4}$ instead of the $\sim r^{-6}$ behavior of the usual Schwarzschild solution.
It would be interesting to understand to what extent this appealing behavior is a generic phenomenon in other theories. One might wonder, for example, if certain theories  (possibly of order higher than ECG) allow for a complete removal of the black hole singularity --- as expected in a UV-complete theory of gravity.

Another remarkable property of the solutions constructed here is the existence of neutral stable small black holes. Indeed, for a given temperature $T<T_{\rm{max}}$, we have found two possible solutions: a large black hole with $C<0$, and a small one with $C>0$. 
As we explain in appendix \ref{remnants}, these black holes never evaporate completely, which gives rise to long-lived remnants. In the general case of a charged, asymptotically (A)dS black hole, we have also obtained exact formulas for all the relevant thermodynamical quantities. However, a detailed analysis of the physical consequences of these new thermodynamic relations is still lacking, and should be carried out elsewhere.

Let us also mention possible generalizations of the solutions presented here. We constructed black holes with spherical horizons, but it should be easy to extend our solutions to different horizon topologies. This possibility has been already considered for the uncharged case in \cite{Hennigar:2016gkm}. Coupling to more complicated matter fields also seems possible. One could of course try to construct non-static or non-spherically symmetric solutions, although that looks more challenging at first sight. Apart from these aspects, we would like to stress again that the properties of ECG make it quite appealing for holographic applications --- see \eg \cite{Dey:2016pei} for a first approach.

\subsection{On the uniqueness of $D=4$ ECG black holes }
According to our computations, up to cubic order in curvature ECG is the most general four-dimensional higher-order gravity which allows for non-trivial single-function generalizations of Schwarzschild- and RN-(A)dS which reduce to the usual Einstein gravity solutions when the corresponding higher-order couplings are set to zero. This requires some further clarification. In fact, four-dimensional black holes with a single function have  been previously constructed for different higher-order gravities. However, all the known cases fall within one of the following three classes: 
\begin{enumerate}
\item They are ``trivial'' embeddings of the usual Einstein black holes \footnote{Besides, it is not clear what the physical meaning of these embeddings is as, in general, the corresponding solutions do not correspond to the exterior gravitational field of any source.}, \ie the metric of the solutions is exactly the same as for Einstein gravity --- see \eg \cite{delaCruzDombriz:2009et} for $f(R)$  or \cite{Lu:2012xu,Smolic:2013gz} for quadratic gravities.

\item  They are solutions to pure higher-order gravities, \ie the action does not incorporate the Einstein-Hilbert term. For example, pure Weyl-squared gravity --- whose Lagrangian reads $\mathcal{L}=\alpha C_{abcd}C^{abcd}$ --- allows for solutions in four dimensions with a single function \cite{Riegert:1984zz,Klemm:1998kf} but $\mathcal{L}=-2\Lambda_0+R+\alpha C_{abcd}C^{abcd}$ does not \cite{Lu:2012xu}. See also \eg \cite{Oliva:2011xu,Oliva:2010zd}.

\item  They require the fine-tuning of some of the higher-order couplings appearing in the action, so the Einstein gravity limit does not exist. For example, extended single-function solutions to a theory of the form $\mathcal{L}=-(-4\Lambda_0+R)^2/(8\Lambda_0)=-2\Lambda_0+R-R^2/(8\Lambda_0)$, which is a perfect square, can be constructed by simply setting $R=4\Lambda_0$. Examples of this kind of constructions can be found \eg in \cite{Cai:2009ac,Love}.
\end{enumerate}
Our claim on the uniqueness of ECG four-dimensional black holes amongst quadratic and cubic gravities applies instead to the situation that we consider most natural. By ``natural'' we mean the following. Consider a theory extending the general relativity action through
\begin{equation}\label{Exx}
S=\frac{1}{16 \pi G}\int_{\mathcal{M}}d^4x\sqrt{|g|}\left[-2\Lambda_0+R+\sum_i \alpha_i X_i \right]\, ,
\end{equation}
where the $X_i=X_i(R_{abcd},g^{ef})$ are higher-curvature invariants and the $\alpha_i$ are independent parameters. Then, we consider black hole solutions of \req{Exx} which non-trivially extend the Einstein gravity ones and reduce to them as we set $\alpha_i=0$ --- a value for which \req{Exx} is also asked to reduce to the Einstein gravity action. This is naturally the kind of scenario that one expects from an effective-action perspective. Single-function examples satisfying these conditions are known in $D\geq 5$ \eg for Lovelock theories \cite{Boulware:1985wk,Cai:2001dz,Dehghani:2009zzb,deBoer:2009gx} or Quasi-topological gravity \cite{Quasi,Quasi2,Dehghani:2011vu}. We claim that in $D=4$, ECG is the only theory that admits this kind of genuine single-function extensions up to cubic order in curvature. According to our analysis, any other term will either imply $N^{\prime}(r)\neq 0$, or keep the Schwarzschild solution unaffected --- or the condition $N^{\prime}(r)=0$ would be achieved by fine-tuning some of the couplings like in the case discussed in item 3 above. This is a remarkable property of ECG, even more so if we take into account that the motivation for constructing this theory was quite different --- namely the fact that it is the most general dimension-independent quadratic or cubic theory whose linearized spectrum coincides with Einstein's \cite{PabloPablo}. In fact, there seems to be a connection between theories which only propagate a massless graviton in the vacuum, and those which allow for single-function black holes. Examples include again Lovelock \cite{Lovelock1,Lovelock2} and Quasi-topological gravity \cite{Quasi,Quasi2}. But note that the connection between these two different aspects of higher-order theories cannot be an \emph{if and only if} because we know examples of theories with the same linearized spectrum as Einstein gravity which do not posses single-function static and spherically symmetric solutions. This is the case for example of certain $f($Lovelock$)$ theories considered in \cite{Love,Bueno:2016ypa,Karasu:2016ifk}, or even of ECG itself when considered in dimensions higher than four --- see the next subsection.

\subsection{On higher-dimensional ECG black holes}
Let us close the paper by mentioning that the generalization of the solutions presented here to higher-dimensions is not straightforward. As we have seen, in $D=4$ we were able to set $N=1$, which allowed us to reduce the problem to a second-order differential equation for $f$. However, this property of four-dimensional ECG no longer holds for $D=5$ since in that case two independent functions are required instead --- the same presumably holding for $D> 5$. Hence, in dimensions higher than four, the construction of solutions should be considerably more involved.


\label{discussion}

\begin{acknowledgments} 
We are thankful to Robert Mann, Rob Myers, Julio Oliva, Tom\'as Ort\'in and C. S. Shahbazi for useful comments. The work of PB was supported by a postdoctoral fellowship from the Fund for Scientific Research - Flanders (FWO). PB also acknowledges support from the Delta ITP Visitors Programme. The work of PAC was supported by a ``la Caixa-Severo Ochoa'' International pre-doctoral grant and in part by the Spanish Ministry of Science and Education grants FPA2012-35043-C02-01 and FPA2015-66793-P and the Centro de Excelencia Severo Ochoa
Program grant SEV-2012-0249.\end{acknowledgments}

\appendix

\section{Black hole remnants?}\label{remnants}
In this appendix we focus again on the asymptotically flat solutions described in sections \ref{asy} and \ref{them}. Our goal here is to describe an interesting feature of these solutions. This is the fact that, as opposed to Schwarzschild, the ECG solutions can describe long-lived --- eternal for all practical purposes --- microscopic black holes, without affecting the usual Einstein gravity physics at astrophysical scales.

Black holes emit a black-body radiation at the Hawking temperature $T$ \cite{Hawking:1974sw}. As a consequence, as seen by an asymptotic observer, they lose mass at a rate determined by the Stefan-Boltzmann law,
\begin{equation}
\frac{dM}{dt}=-P=- 4\pi r_h^2\,  \sigma\cdot  T^4\, , \quad  \text{where} \quad \sigma= \frac{\pi^2}{60}
\label{power}
\end{equation}
is the Stefan-Boltzmann constant. In the case of a Schwarzschild black hole, the total power emitted is greater for lower masses, so the rate at which the mass is lost increases as the mass becomes smaller. As a consequence, the black hole evaporates after a finite time. However, for the ECG black holes constructed in sections  \ref{asy} and \ref{them}, the situation is different because its temperature vanishes for lower masses. Indeed,  when $GM<40/49\cdot (27\lambda G^2 /7)^{1/4}$, the power emitted starts to decrease with the mass. Therefore,  when $GM\ll(G^{2}\lambda)^{1/4}$ one is left with a remnant which never evaporates completely. We can compute explicitly the half-life of such remnant. For $GM\ll (G^{2}\lambda)^{1/4}$, we get approximately  $r_h\approx (6\sqrt{3\lambda  }G^2 M)^{1/3}$ and $T\approx r_h / (2\pi\sqrt{48G^2\lambda})$. Then we can integrate (\ref{power}) to get
\begin{equation}
M(t)=\frac{M_0}{1+t/t_0}\, , \quad \text{where} \quad t_0=\frac{5120\pi \lambda}{ M_0}
\end{equation}
is the half-life of the remnant, \ie the time that it takes to lose half of its mass.

Quick estimations of $t_0$ compatible with imposing that macroscopic physics is undistinguishable from Einstein gravity --- \ie $(G^2\lambda)^{1/4}/G\mathsf{M}\ll 1$ for masses $\mathsf{M}$ of the order of astrophysical objects --- show that ECG indeed allows for long-lived black-hole remnants.  For example, if we impose $(G^2\lambda)\sim m_{\rm proton}^{-4}$ and an initial mass $M_0\sim  m_{\rm proton}$ --- which fulfills the remnant condition $GM_0\ll (G^2\lambda)^{1/4}$ --- the half-life would be $t_0\sim 10^{38}\,\times$ the age of the universe. Note that the condition $(G^2\lambda)^{1/4}/G\mathsf{M}\ll 1$ is comfortably satisfied by this value of $G^2\lambda$ for example if we set $\mathsf{M}\sim \mathsf{M}_{\odot}$. In that case, it translates into $G^2 \lambda\ll 10^{64}\cdot m_{\rm proton}^{-4}$, which is of course satisfied by $(G^2\lambda)\sim m_{\rm proton}^{-4}$. 

The point of this simple computation is to stress that it is in principle possible to conceive higher-curvature extensions of Einstein gravity which allow for long-lived microscopic black holes, but do not affect the usual Einstein-gravity physics at astrophysical or cosmological scales. We find this feature quite appealing and worth further study. In particular, it would be interesting to understand how generically the above features occur when the Einstein-Hilbert action is supplemented by higher-order terms. The existence of black-hole remnants could have important observational --- and theoretical --- implications. Note that observations similar to the ones reported in this appendix have been made elsewhere for other classes of higher-order theories --- see \eg \cite{Olmo:2013gqa,Lobo:2013prg,Chen:2014jwq} and references therein.


\bibliography{Gravities}

 \newcommand{\noop}[1]{}
\begin{thebibliography}{44}%
\makeatletter
\providecommand \@ifxundefined [1]{%
 \@ifx{#1\undefined}
}%
\providecommand \@ifnum [1]{%
 \ifnum #1\expandafter \@firstoftwo
 \else \expandafter \@secondoftwo
 \fi
}%
\providecommand \@ifx [1]{%
 \ifx #1\expandafter \@firstoftwo
 \else \expandafter \@secondoftwo
 \fi
}%
\providecommand \natexlab [1]{#1}%
\providecommand \enquote  [1]{``#1''}%
\providecommand \bibnamefont  [1]{#1}%
\providecommand \bibfnamefont [1]{#1}%
\providecommand \citenamefont [1]{#1}%
\providecommand \href@noop [0]{\@secondoftwo}%
\providecommand \href [0]{\begingroup \@sanitize@url \@href}%
\providecommand \@href[1]{\@@startlink{#1}\@@href}%
\providecommand \@@href[1]{\endgroup#1\@@endlink}%
\providecommand \@sanitize@url [0]{\catcode `\\12\catcode `\$12\catcode
  `\&12\catcode `\#12\catcode `\^12\catcode `\_12\catcode `\%12\relax}%
\providecommand \@@startlink[1]{}%
\providecommand \@@endlink[0]{}%
\providecommand \url  [0]{\begingroup\@sanitize@url \@url }%
\providecommand \@url [1]{\endgroup\@href {#1}{\urlprefix }}%
\providecommand \urlprefix  [0]{URL }%
\providecommand \Eprint [0]{\href }%
\providecommand \doibase [0]{http://dx.doi.org/}%
\providecommand \selectlanguage [0]{\@gobble}%
\providecommand \bibinfo  [0]{\@secondoftwo}%
\providecommand \bibfield  [0]{\@secondoftwo}%
\providecommand \translation [1]{[#1]}%
\providecommand \BibitemOpen [0]{}%
\providecommand \bibitemStop [0]{}%
\providecommand \bibitemNoStop [0]{.\EOS\space}%
\providecommand \EOS [0]{\spacefactor3000\relax}%
\providecommand \BibitemShut  [1]{\csname bibitem#1\endcsname}%
\let\auto@bib@innerbib\@empty
\bibitem [{\citenamefont {Bueno}\ and\ \citenamefont
  {Cano}(2016)}]{PabloPablo}%
  \BibitemOpen
  \bibfield  {author} {\bibinfo {author} {\bibfnamefont {P.}~\bibnamefont
  {Bueno}}\ and\ \bibinfo {author} {\bibfnamefont {P.~A.}\ \bibnamefont
  {Cano}},\ }\href@noop {} {\  (\bibinfo {year} {2016})},\ \Eprint
  {http://arxiv.org/abs/1607.06463} {arXiv:1607.06463 [hep-th]} \BibitemShut
  {NoStop}%
\bibitem [{\citenamefont {Lovelock}(1970)}]{Lovelock1}%
  \BibitemOpen
  \bibfield  {author} {\bibinfo {author} {\bibfnamefont {D.}~\bibnamefont
  {Lovelock}},\ }\href {\doibase 10.1007/BF01817753} {\bibfield  {journal}
  {\bibinfo  {journal} {aequationes mathematicae}\ }\textbf {\bibinfo {volume}
  {4}},\ \bibinfo {pages} {127} (\bibinfo {year} {1970})}\BibitemShut {NoStop}%
\bibitem [{\citenamefont {Lovelock}(1971)}]{Lovelock2}%
  \BibitemOpen
  \bibfield  {author} {\bibinfo {author} {\bibfnamefont {D.}~\bibnamefont
  {Lovelock}},\ }\href {\doibase 10.1063/1.1665613} {\bibfield  {journal}
  {\bibinfo  {journal} {J. Math. Phys.}\ }\textbf {\bibinfo {volume} {12}},\
  \bibinfo {pages} {498} (\bibinfo {year} {1971})}\BibitemShut {NoStop}%
\bibitem [{\citenamefont {Wald}(1993)}]{Wald:1993nt}%
  \BibitemOpen
  \bibfield  {author} {\bibinfo {author} {\bibfnamefont {R.~M.}\ \bibnamefont
  {Wald}},\ }\href {\doibase 10.1103/PhysRevD.48.R3427} {\bibfield  {journal}
  {\bibinfo  {journal} {Phys. Rev.}\ }\textbf {\bibinfo {volume} {D48}},\
  \bibinfo {pages} {3427} (\bibinfo {year} {1993})},\ \Eprint
  {http://arxiv.org/abs/gr-qc/9307038} {arXiv:gr-qc/9307038 [gr-qc]}
  \BibitemShut {NoStop}%
\bibitem [{\citenamefont {Hennigar}\ and\ \citenamefont
  {Mann}(2016)}]{Hennigar:2016gkm}%
  \BibitemOpen
  \bibfield  {author} {\bibinfo {author} {\bibfnamefont {R.~A.}\ \bibnamefont
  {Hennigar}}\ and\ \bibinfo {author} {\bibfnamefont {R.~B.}\ \bibnamefont
  {Mann}},\ }\href@noop {} {\  (\bibinfo {year} {2016})},\ \Eprint
  {http://arxiv.org/abs/1610.06675} {arXiv:1610.06675 [hep-th]} \BibitemShut
  {NoStop}%
\bibitem [{Note1()}]{Note1}%
  \BibitemOpen
  \bibinfo {note} {The extension to hyperbolic and planar horizons is
  straightforward.}\BibitemShut {Stop}%
\bibitem [{\citenamefont {Myers}\ and\ \citenamefont {Robinson}(2010)}]{Quasi}%
  \BibitemOpen
  \bibfield  {author} {\bibinfo {author} {\bibfnamefont {R.~C.}\ \bibnamefont
  {Myers}}\ and\ \bibinfo {author} {\bibfnamefont {B.}~\bibnamefont
  {Robinson}},\ }\href {\doibase 10.1007/JHEP08(2010)067} {\bibfield  {journal}
  {\bibinfo  {journal} {JHEP}\ }\textbf {\bibinfo {volume} {08}},\ \bibinfo
  {pages} {067} (\bibinfo {year} {2010})},\ \Eprint
  {http://arxiv.org/abs/1003.5357} {arXiv:1003.5357 [gr-qc]} \BibitemShut
  {NoStop}%
\bibitem [{Note2()}]{Note2}%
  \BibitemOpen
  \bibinfo {note} {Note that $\protect \mathcal {E}_{a b}=\protect \frac
  {1}{\protect \sqrt {|g|}}\protect \frac {\delta S}{\delta g^{ab}}
  $}\BibitemShut {NoStop}%
\bibitem [{\citenamefont {Lu}\ \emph {et~al.}(2015)\citenamefont {Lu},
  \citenamefont {Perkins}, \citenamefont {Pope},\ and\ \citenamefont
  {Stelle}}]{Lu:2015cqa}%
  \BibitemOpen
  \bibfield  {author} {\bibinfo {author} {\bibfnamefont {H.}~\bibnamefont
  {Lu}}, \bibinfo {author} {\bibfnamefont {A.}~\bibnamefont {Perkins}},
  \bibinfo {author} {\bibfnamefont {C.~N.}\ \bibnamefont {Pope}}, \ and\
  \bibinfo {author} {\bibfnamefont {K.~S.}\ \bibnamefont {Stelle}},\ }\href
  {\doibase 10.1103/PhysRevLett.114.171601} {\bibfield  {journal} {\bibinfo
  {journal} {Phys. Rev. Lett.}\ }\textbf {\bibinfo {volume} {114}},\ \bibinfo
  {pages} {171601} (\bibinfo {year} {2015})},\ \Eprint
  {http://arxiv.org/abs/1502.01028} {arXiv:1502.01028 [hep-th]} \BibitemShut
  {NoStop}%
\bibitem [{\citenamefont {Lü}\ \emph {et~al.}(2015)\citenamefont {Lü},
  \citenamefont {Perkins}, \citenamefont {Pope},\ and\ \citenamefont
  {Stelle}}]{Lu:2015psa}%
  \BibitemOpen
  \bibfield  {author} {\bibinfo {author} {\bibfnamefont {H.}~\bibnamefont
  {Lü}}, \bibinfo {author} {\bibfnamefont {A.}~\bibnamefont {Perkins}},
  \bibinfo {author} {\bibfnamefont {C.~N.}\ \bibnamefont {Pope}}, \ and\
  \bibinfo {author} {\bibfnamefont {K.~S.}\ \bibnamefont {Stelle}},\ }\href
  {\doibase 10.1103/PhysRevD.92.124019} {\bibfield  {journal} {\bibinfo
  {journal} {Phys. Rev.}\ }\textbf {\bibinfo {volume} {D92}},\ \bibinfo {pages}
  {124019} (\bibinfo {year} {2015})},\ \Eprint
  {http://arxiv.org/abs/1508.00010} {arXiv:1508.00010 [hep-th]} \BibitemShut
  {NoStop}%
\bibitem [{\citenamefont {Bueno}\ \emph
  {et~al.}(2016{\natexlab{a}})\citenamefont {Bueno}, \citenamefont {Cano},
  \citenamefont {Min},\ and\ \citenamefont {Visser}}]{Bueno:2016ypa}%
  \BibitemOpen
  \bibfield  {author} {\bibinfo {author} {\bibfnamefont {P.}~\bibnamefont
  {Bueno}}, \bibinfo {author} {\bibfnamefont {P.~A.}\ \bibnamefont {Cano}},
  \bibinfo {author} {\bibfnamefont {V.~S.}\ \bibnamefont {Min}}, \ and\
  \bibinfo {author} {\bibfnamefont {M.~R.}\ \bibnamefont {Visser}},\
  }\href@noop {} {\  (\bibinfo {year} {2016}{\natexlab{a}})},\ \Eprint
  {http://arxiv.org/abs/1610.08519} {arXiv:1610.08519 [hep-th]} \BibitemShut
  {NoStop}%
\bibitem [{\citenamefont {Abbott}\ and\ \citenamefont
  {Deser}(1982)}]{Abbott:1981ff}%
  \BibitemOpen
  \bibfield  {author} {\bibinfo {author} {\bibfnamefont {L.~F.}\ \bibnamefont
  {Abbott}}\ and\ \bibinfo {author} {\bibfnamefont {S.}~\bibnamefont {Deser}},\
  }\href {\doibase 10.1016/0550-3213(82)90049-9} {\bibfield  {journal}
  {\bibinfo  {journal} {Nucl. Phys.}\ }\textbf {\bibinfo {volume} {B195}},\
  \bibinfo {pages} {76} (\bibinfo {year} {1982})}\BibitemShut {NoStop}%
\bibitem [{\citenamefont {Deser}\ and\ \citenamefont
  {Tekin}(2003)}]{Deser:2002jk}%
  \BibitemOpen
  \bibfield  {author} {\bibinfo {author} {\bibfnamefont {S.}~\bibnamefont
  {Deser}}\ and\ \bibinfo {author} {\bibfnamefont {B.}~\bibnamefont {Tekin}},\
  }\href {\doibase 10.1103/PhysRevD.67.084009} {\bibfield  {journal} {\bibinfo
  {journal} {Phys. Rev.}\ }\textbf {\bibinfo {volume} {D67}},\ \bibinfo {pages}
  {084009} (\bibinfo {year} {2003})},\ \Eprint
  {http://arxiv.org/abs/hep-th/0212292} {arXiv:hep-th/0212292 [hep-th]}
  \BibitemShut {NoStop}%
\bibitem [{Note3()}]{Note3}%
  \BibitemOpen
  \bibinfo {note} {As we see from (\ref {fequation}), at the horizon ---
  {\protect \it i.e.,}\ when $f=0$ --- the term which multiplies $f''$
  vanishes. This can give rise to non-differentiability on the horizon of some
  of the solutions, so imposing that the horizon is regular is indeed a strong
  restriction.}\BibitemShut {Stop}%
\bibitem [{Note4()}]{Note4}%
  \BibitemOpen
  \bibinfo {note} {In fact there is another solution with a minus sign in front
  of the square root, but that choice does not reproduce the correct limit when
  $\lambda =0$.}\BibitemShut {Stop}%
\bibitem [{\citenamefont {Hawking}(1975)}]{Hawking:1974sw}%
  \BibitemOpen
  \bibfield  {author} {\bibinfo {author} {\bibfnamefont {S.~W.}\ \bibnamefont
  {Hawking}},\ }\bibfield  {booktitle} {\emph {\bibinfo {booktitle} {{In
  *Gibbons, G.W. (ed.), Hawking, S.W. (ed.): Euclidean quantum gravity*
  167-188}}},\ }\href {\doibase 10.1007/BF02345020} {\bibfield  {journal}
  {\bibinfo  {journal} {Commun. Math. Phys.}\ }\textbf {\bibinfo {volume}
  {43}},\ \bibinfo {pages} {199} (\bibinfo {year} {1975})},\ \bibinfo {note}
  {[,167(1975)]}\BibitemShut {NoStop}%
\bibitem [{\citenamefont {Bardeen}\ \emph {et~al.}(1973)\citenamefont
  {Bardeen}, \citenamefont {Carter},\ and\ \citenamefont
  {Hawking}}]{Bardeen:1973gs}%
  \BibitemOpen
  \bibfield  {author} {\bibinfo {author} {\bibfnamefont {J.~M.}\ \bibnamefont
  {Bardeen}}, \bibinfo {author} {\bibfnamefont {B.}~\bibnamefont {Carter}}, \
  and\ \bibinfo {author} {\bibfnamefont {S.~W.}\ \bibnamefont {Hawking}},\
  }\href {\doibase 10.1007/BF01645742} {\bibfield  {journal} {\bibinfo
  {journal} {Commun. Math. Phys.}\ }\textbf {\bibinfo {volume} {31}},\ \bibinfo
  {pages} {161} (\bibinfo {year} {1973})}\BibitemShut {NoStop}%
\bibitem [{\citenamefont {Bekenstein}(1973)}]{Bekenstein:1973ur}%
  \BibitemOpen
  \bibfield  {author} {\bibinfo {author} {\bibfnamefont {J.~D.}\ \bibnamefont
  {Bekenstein}},\ }\href {\doibase 10.1103/PhysRevD.7.2333} {\bibfield
  {journal} {\bibinfo  {journal} {Phys. Rev.}\ }\textbf {\bibinfo {volume}
  {D7}},\ \bibinfo {pages} {2333} (\bibinfo {year} {1973})}\BibitemShut
  {NoStop}%
\bibitem [{\citenamefont {Bekenstein}(1974)}]{Bekenstein:1974ax}%
  \BibitemOpen
  \bibfield  {author} {\bibinfo {author} {\bibfnamefont {J.~D.}\ \bibnamefont
  {Bekenstein}},\ }\href {\doibase 10.1103/PhysRevD.9.3292} {\bibfield
  {journal} {\bibinfo  {journal} {Phys. Rev.}\ }\textbf {\bibinfo {volume}
  {D9}},\ \bibinfo {pages} {3292} (\bibinfo {year} {1974})}\BibitemShut
  {NoStop}%
\bibitem [{\citenamefont {Iyer}\ and\ \citenamefont
  {Wald}(1994)}]{Iyer:1994ys}%
  \BibitemOpen
  \bibfield  {author} {\bibinfo {author} {\bibfnamefont {V.}~\bibnamefont
  {Iyer}}\ and\ \bibinfo {author} {\bibfnamefont {R.~M.}\ \bibnamefont
  {Wald}},\ }\href {\doibase 10.1103/PhysRevD.50.846} {\bibfield  {journal}
  {\bibinfo  {journal} {Phys. Rev.}\ }\textbf {\bibinfo {volume} {D50}},\
  \bibinfo {pages} {846} (\bibinfo {year} {1994})},\ \Eprint
  {http://arxiv.org/abs/gr-qc/9403028} {arXiv:gr-qc/9403028 [gr-qc]}
  \BibitemShut {NoStop}%
\bibitem [{\citenamefont {Jacobson}\ \emph {et~al.}(1994)\citenamefont
  {Jacobson}, \citenamefont {Kang},\ and\ \citenamefont
  {Myers}}]{Jacobson:1993vj}%
  \BibitemOpen
  \bibfield  {author} {\bibinfo {author} {\bibfnamefont {T.}~\bibnamefont
  {Jacobson}}, \bibinfo {author} {\bibfnamefont {G.}~\bibnamefont {Kang}}, \
  and\ \bibinfo {author} {\bibfnamefont {R.~C.}\ \bibnamefont {Myers}},\ }\href
  {\doibase 10.1103/PhysRevD.49.6587} {\bibfield  {journal} {\bibinfo
  {journal} {Phys. Rev.}\ }\textbf {\bibinfo {volume} {D49}},\ \bibinfo {pages}
  {6587} (\bibinfo {year} {1994})},\ \Eprint
  {http://arxiv.org/abs/gr-qc/9312023} {arXiv:gr-qc/9312023 [gr-qc]}
  \BibitemShut {NoStop}%
\bibitem [{\citenamefont {Ortin}(2004)}]{Ortin:2004ms}%
  \BibitemOpen
  \bibfield  {author} {\bibinfo {author} {\bibfnamefont {T.}~\bibnamefont
  {Ortin}},\ }\href
  {http://www.cambridge.org/uk/catalogue/catalogue.asp?isbn=0521824753} {\emph
  {\bibinfo {title} {{Gravity and strings}}}}\ (\bibinfo  {publisher}
  {Cambridge Univ. Press},\ \bibinfo {year} {2004})\BibitemShut {NoStop}%
\bibitem [{\citenamefont {Myers}\ and\ \citenamefont
  {Simon}(1988)}]{Myers:1988ze}%
  \BibitemOpen
  \bibfield  {author} {\bibinfo {author} {\bibfnamefont {R.~C.}\ \bibnamefont
  {Myers}}\ and\ \bibinfo {author} {\bibfnamefont {J.~Z.}\ \bibnamefont
  {Simon}},\ }\href {\doibase 10.1103/PhysRevD.38.2434} {\bibfield  {journal}
  {\bibinfo  {journal} {Phys. Rev.}\ }\textbf {\bibinfo {volume} {D38}},\
  \bibinfo {pages} {2434} (\bibinfo {year} {1988})}\BibitemShut {NoStop}%
\bibitem [{\citenamefont {Dey}\ \emph {et~al.}(2016)\citenamefont {Dey},
  \citenamefont {Roy},\ and\ \citenamefont {Sarkar}}]{Dey:2016pei}%
  \BibitemOpen
  \bibfield  {author} {\bibinfo {author} {\bibfnamefont {A.}~\bibnamefont
  {Dey}}, \bibinfo {author} {\bibfnamefont {P.}~\bibnamefont {Roy}}, \ and\
  \bibinfo {author} {\bibfnamefont {T.}~\bibnamefont {Sarkar}},\ }\href@noop {}
  {\  (\bibinfo {year} {2016})},\ \Eprint {http://arxiv.org/abs/1609.02290}
  {arXiv:1609.02290 [hep-th]} \BibitemShut {NoStop}%
\bibitem [{Note5()}]{Note5}%
  \BibitemOpen
  \bibinfo {note} {Besides, it is not clear what the physical meaning of these
  embeddings is as, in general, the corresponding solutions do not correspond
  to the exterior gravitational field of any source.}\BibitemShut {Stop}%
\bibitem [{\citenamefont {de~la Cruz-Dombriz}\ \emph
  {et~al.}(2009)\citenamefont {de~la Cruz-Dombriz}, \citenamefont {Dobado},\
  and\ \citenamefont {Maroto}}]{delaCruzDombriz:2009et}%
  \BibitemOpen
  \bibfield  {author} {\bibinfo {author} {\bibfnamefont {A.}~\bibnamefont
  {de~la Cruz-Dombriz}}, \bibinfo {author} {\bibfnamefont {A.}~\bibnamefont
  {Dobado}}, \ and\ \bibinfo {author} {\bibfnamefont {A.~L.}\ \bibnamefont
  {Maroto}},\ }\href {\doibase 10.1103/PhysRevD.83.029903,
  10.1103/PhysRevD.80.124011} {\bibfield  {journal} {\bibinfo  {journal} {Phys.
  Rev.}\ }\textbf {\bibinfo {volume} {D80}},\ \bibinfo {pages} {124011}
  (\bibinfo {year} {2009})},\ \bibinfo {note} {[Erratum: Phys.
  Rev.D83,029903(2011)]},\ \Eprint {http://arxiv.org/abs/0907.3872}
  {arXiv:0907.3872 [gr-qc]} \BibitemShut {NoStop}%
\bibitem [{\citenamefont {Lu}\ \emph {et~al.}(2012)\citenamefont {Lu},
  \citenamefont {Pang}, \citenamefont {Pope},\ and\ \citenamefont
  {Vazquez-Poritz}}]{Lu:2012xu}%
  \BibitemOpen
  \bibfield  {author} {\bibinfo {author} {\bibfnamefont {H.}~\bibnamefont
  {Lu}}, \bibinfo {author} {\bibfnamefont {Y.}~\bibnamefont {Pang}}, \bibinfo
  {author} {\bibfnamefont {C.~N.}\ \bibnamefont {Pope}}, \ and\ \bibinfo
  {author} {\bibfnamefont {J.~F.}\ \bibnamefont {Vazquez-Poritz}},\ }\href
  {\doibase 10.1103/PhysRevD.86.044011} {\bibfield  {journal} {\bibinfo
  {journal} {Phys. Rev.}\ }\textbf {\bibinfo {volume} {D86}},\ \bibinfo {pages}
  {044011} (\bibinfo {year} {2012})},\ \Eprint {http://arxiv.org/abs/1204.1062}
  {arXiv:1204.1062 [hep-th]} \BibitemShut {NoStop}%
\bibitem [{\citenamefont {Smolic}\ and\ \citenamefont
  {Taylor}(2013)}]{Smolic:2013gz}%
  \BibitemOpen
  \bibfield  {author} {\bibinfo {author} {\bibfnamefont {J.}~\bibnamefont
  {Smolic}}\ and\ \bibinfo {author} {\bibfnamefont {M.}~\bibnamefont
  {Taylor}},\ }\href {\doibase 10.1007/JHEP06(2013)096} {\bibfield  {journal}
  {\bibinfo  {journal} {JHEP}\ }\textbf {\bibinfo {volume} {06}},\ \bibinfo
  {pages} {096} (\bibinfo {year} {2013})},\ \Eprint
  {http://arxiv.org/abs/1301.5205} {arXiv:1301.5205 [hep-th]} \BibitemShut
  {NoStop}%
\bibitem [{\citenamefont {Riegert}(1984)}]{Riegert:1984zz}%
  \BibitemOpen
  \bibfield  {author} {\bibinfo {author} {\bibfnamefont {R.~J.}\ \bibnamefont
  {Riegert}},\ }\href {\doibase 10.1103/PhysRevLett.53.315} {\bibfield
  {journal} {\bibinfo  {journal} {Phys. Rev. Lett.}\ }\textbf {\bibinfo
  {volume} {53}},\ \bibinfo {pages} {315} (\bibinfo {year} {1984})}\BibitemShut
  {NoStop}%
\bibitem [{\citenamefont {Klemm}(1998)}]{Klemm:1998kf}%
  \BibitemOpen
  \bibfield  {author} {\bibinfo {author} {\bibfnamefont {D.}~\bibnamefont
  {Klemm}},\ }\href {\doibase 10.1088/0264-9381/15/10/020} {\bibfield
  {journal} {\bibinfo  {journal} {Class. Quant. Grav.}\ }\textbf {\bibinfo
  {volume} {15}},\ \bibinfo {pages} {3195} (\bibinfo {year} {1998})},\ \Eprint
  {http://arxiv.org/abs/gr-qc/9808051} {arXiv:gr-qc/9808051 [gr-qc]}
  \BibitemShut {NoStop}%
\bibitem [{\citenamefont {Oliva}\ and\ \citenamefont
  {Ray}(2011)}]{Oliva:2011xu}%
  \BibitemOpen
  \bibfield  {author} {\bibinfo {author} {\bibfnamefont {J.}~\bibnamefont
  {Oliva}}\ and\ \bibinfo {author} {\bibfnamefont {S.}~\bibnamefont {Ray}},\
  }\href {\doibase 10.1088/0264-9381/28/17/175007} {\bibfield  {journal}
  {\bibinfo  {journal} {Class. Quant. Grav.}\ }\textbf {\bibinfo {volume}
  {28}},\ \bibinfo {pages} {175007} (\bibinfo {year} {2011})},\ \Eprint
  {http://arxiv.org/abs/1104.1205} {arXiv:1104.1205 [gr-qc]} \BibitemShut
  {NoStop}%
\bibitem [{\citenamefont {Oliva}\ and\ \citenamefont
  {Ray}(2010{\natexlab{a}})}]{Oliva:2010zd}%
  \BibitemOpen
  \bibfield  {author} {\bibinfo {author} {\bibfnamefont {J.}~\bibnamefont
  {Oliva}}\ and\ \bibinfo {author} {\bibfnamefont {S.}~\bibnamefont {Ray}},\
  }\href {\doibase 10.1103/PhysRevD.82.124030} {\bibfield  {journal} {\bibinfo
  {journal} {Phys. Rev.}\ }\textbf {\bibinfo {volume} {D82}},\ \bibinfo {pages}
  {124030} (\bibinfo {year} {2010}{\natexlab{a}})},\ \Eprint
  {http://arxiv.org/abs/1004.0737} {arXiv:1004.0737 [gr-qc]} \BibitemShut
  {NoStop}%
\bibitem [{\citenamefont {Cai}\ \emph {et~al.}(2009)\citenamefont {Cai},
  \citenamefont {Liu},\ and\ \citenamefont {Sun}}]{Cai:2009ac}%
  \BibitemOpen
  \bibfield  {author} {\bibinfo {author} {\bibfnamefont {R.-G.}\ \bibnamefont
  {Cai}}, \bibinfo {author} {\bibfnamefont {Y.}~\bibnamefont {Liu}}, \ and\
  \bibinfo {author} {\bibfnamefont {Y.-W.}\ \bibnamefont {Sun}},\ }\href
  {\doibase 10.1088/1126-6708/2009/10/080} {\bibfield  {journal} {\bibinfo
  {journal} {JHEP}\ }\textbf {\bibinfo {volume} {10}},\ \bibinfo {pages} {080}
  (\bibinfo {year} {2009})},\ \Eprint {http://arxiv.org/abs/0909.2807}
  {arXiv:0909.2807 [hep-th]} \BibitemShut {NoStop}%
\bibitem [{\citenamefont {Bueno}\ \emph
  {et~al.}(2016{\natexlab{b}})\citenamefont {Bueno}, \citenamefont {Cano},
  \citenamefont {Lasso},\ and\ \citenamefont {Ramírez}}]{Love}%
  \BibitemOpen
  \bibfield  {author} {\bibinfo {author} {\bibfnamefont {P.}~\bibnamefont
  {Bueno}}, \bibinfo {author} {\bibfnamefont {P.~A.}\ \bibnamefont {Cano}},
  \bibinfo {author} {\bibfnamefont {A.~O.}\ \bibnamefont {Lasso}}, \ and\
  \bibinfo {author} {\bibfnamefont {P.~F.}\ \bibnamefont {Ramírez}},\ }\href
  {\doibase 10.1007/JHEP04(2016)028} {\bibfield  {journal} {\bibinfo  {journal}
  {JHEP}\ }\textbf {\bibinfo {volume} {04}},\ \bibinfo {pages} {028} (\bibinfo
  {year} {2016}{\natexlab{b}})},\ \Eprint {http://arxiv.org/abs/1602.07310}
  {arXiv:1602.07310 [hep-th]} \BibitemShut {NoStop}%
\bibitem [{\citenamefont {Boulware}\ and\ \citenamefont
  {Deser}(1985)}]{Boulware:1985wk}%
  \BibitemOpen
  \bibfield  {author} {\bibinfo {author} {\bibfnamefont {D.~G.}\ \bibnamefont
  {Boulware}}\ and\ \bibinfo {author} {\bibfnamefont {S.}~\bibnamefont
  {Deser}},\ }\href {\doibase 10.1103/PhysRevLett.55.2656} {\bibfield
  {journal} {\bibinfo  {journal} {Phys. Rev. Lett.}\ }\textbf {\bibinfo
  {volume} {55}},\ \bibinfo {pages} {2656} (\bibinfo {year}
  {1985})}\BibitemShut {NoStop}%
\bibitem [{\citenamefont {Cai}(2002)}]{Cai:2001dz}%
  \BibitemOpen
  \bibfield  {author} {\bibinfo {author} {\bibfnamefont {R.-G.}\ \bibnamefont
  {Cai}},\ }\href {\doibase 10.1103/PhysRevD.65.084014} {\bibfield  {journal}
  {\bibinfo  {journal} {Phys. Rev.}\ }\textbf {\bibinfo {volume} {D65}},\
  \bibinfo {pages} {084014} (\bibinfo {year} {2002})},\ \Eprint
  {http://arxiv.org/abs/hep-th/0109133} {arXiv:hep-th/0109133 [hep-th]}
  \BibitemShut {NoStop}%
\bibitem [{\citenamefont {Dehghani}\ and\ \citenamefont
  {Pourhasan}(2009)}]{Dehghani:2009zzb}%
  \BibitemOpen
  \bibfield  {author} {\bibinfo {author} {\bibfnamefont {M.~H.}\ \bibnamefont
  {Dehghani}}\ and\ \bibinfo {author} {\bibfnamefont {R.}~\bibnamefont
  {Pourhasan}},\ }\href {\doibase 10.1103/PhysRevD.79.064015} {\bibfield
  {journal} {\bibinfo  {journal} {Phys. Rev.}\ }\textbf {\bibinfo {volume}
  {D79}},\ \bibinfo {pages} {064015} (\bibinfo {year} {2009})},\ \Eprint
  {http://arxiv.org/abs/0903.4260} {arXiv:0903.4260 [gr-qc]} \BibitemShut
  {NoStop}%
\bibitem [{\citenamefont {de~Boer}\ \emph {et~al.}(2010)\citenamefont
  {de~Boer}, \citenamefont {Kulaxizi},\ and\ \citenamefont
  {Parnachev}}]{deBoer:2009gx}%
  \BibitemOpen
  \bibfield  {author} {\bibinfo {author} {\bibfnamefont {J.}~\bibnamefont
  {de~Boer}}, \bibinfo {author} {\bibfnamefont {M.}~\bibnamefont {Kulaxizi}}, \
  and\ \bibinfo {author} {\bibfnamefont {A.}~\bibnamefont {Parnachev}},\ }\href
  {\doibase 10.1007/JHEP06(2010)008} {\bibfield  {journal} {\bibinfo  {journal}
  {JHEP}\ }\textbf {\bibinfo {volume} {06}},\ \bibinfo {pages} {008} (\bibinfo
  {year} {2010})},\ \Eprint {http://arxiv.org/abs/0912.1877} {arXiv:0912.1877
  [hep-th]} \BibitemShut {NoStop}%
\bibitem [{\citenamefont {Oliva}\ and\ \citenamefont
  {Ray}(2010{\natexlab{b}})}]{Quasi2}%
  \BibitemOpen
  \bibfield  {author} {\bibinfo {author} {\bibfnamefont {J.}~\bibnamefont
  {Oliva}}\ and\ \bibinfo {author} {\bibfnamefont {S.}~\bibnamefont {Ray}},\
  }\href {\doibase 10.1088/0264-9381/27/22/225002} {\bibfield  {journal}
  {\bibinfo  {journal} {Class. Quant. Grav.}\ }\textbf {\bibinfo {volume}
  {27}},\ \bibinfo {pages} {225002} (\bibinfo {year} {2010}{\natexlab{b}})},\
  \Eprint {http://arxiv.org/abs/1003.4773} {arXiv:1003.4773 [gr-qc]}
  \BibitemShut {NoStop}%
\bibitem [{\citenamefont {Dehghani}\ \emph {et~al.}(2012)\citenamefont
  {Dehghani}, \citenamefont {Bazrafshan}, \citenamefont {Mann}, \citenamefont
  {Mehdizadeh}, \citenamefont {Ghanaatian},\ and\ \citenamefont
  {Vahidinia}}]{Dehghani:2011vu}%
  \BibitemOpen
  \bibfield  {author} {\bibinfo {author} {\bibfnamefont {M.~H.}\ \bibnamefont
  {Dehghani}}, \bibinfo {author} {\bibfnamefont {A.}~\bibnamefont
  {Bazrafshan}}, \bibinfo {author} {\bibfnamefont {R.~B.}\ \bibnamefont
  {Mann}}, \bibinfo {author} {\bibfnamefont {M.~R.}\ \bibnamefont
  {Mehdizadeh}}, \bibinfo {author} {\bibfnamefont {M.}~\bibnamefont
  {Ghanaatian}}, \ and\ \bibinfo {author} {\bibfnamefont {M.~H.}\ \bibnamefont
  {Vahidinia}},\ }\href {\doibase 10.1103/PhysRevD.85.104009} {\bibfield
  {journal} {\bibinfo  {journal} {Phys. Rev.}\ }\textbf {\bibinfo {volume}
  {D85}},\ \bibinfo {pages} {104009} (\bibinfo {year} {2012})},\ \Eprint
  {http://arxiv.org/abs/1109.4708} {arXiv:1109.4708 [hep-th]} \BibitemShut
  {NoStop}%
\bibitem [{\citenamefont {Karasu}\ \emph {et~al.}(2016)\citenamefont {Karasu},
  \citenamefont {Kenar},\ and\ \citenamefont {Tekin}}]{Karasu:2016ifk}%
  \BibitemOpen
  \bibfield  {author} {\bibinfo {author} {\bibfnamefont {A.}~\bibnamefont
  {Karasu}}, \bibinfo {author} {\bibfnamefont {E.}~\bibnamefont {Kenar}}, \
  and\ \bibinfo {author} {\bibfnamefont {B.}~\bibnamefont {Tekin}},\ }\href
  {\doibase 10.1103/PhysRevD.93.084040} {\bibfield  {journal} {\bibinfo
  {journal} {Phys. Rev.}\ }\textbf {\bibinfo {volume} {D93}},\ \bibinfo {pages}
  {084040} (\bibinfo {year} {2016})},\ \Eprint
  {http://arxiv.org/abs/1602.02567} {arXiv:1602.02567 [hep-th]} \BibitemShut
  {NoStop}%
\bibitem [{\citenamefont {Olmo}\ \emph {et~al.}(2014)\citenamefont {Olmo},
  \citenamefont {Rubiera-Garcia},\ and\ \citenamefont
  {Sanchis-Alepuz}}]{Olmo:2013gqa}%
  \BibitemOpen
  \bibfield  {author} {\bibinfo {author} {\bibfnamefont {G.~J.}\ \bibnamefont
  {Olmo}}, \bibinfo {author} {\bibfnamefont {D.}~\bibnamefont
  {Rubiera-Garcia}}, \ and\ \bibinfo {author} {\bibfnamefont {H.}~\bibnamefont
  {Sanchis-Alepuz}},\ }\href {\doibase 10.1140/epjc/s10052-014-2804-8}
  {\bibfield  {journal} {\bibinfo  {journal} {Eur. Phys. J.}\ }\textbf
  {\bibinfo {volume} {C74}},\ \bibinfo {pages} {2804} (\bibinfo {year}
  {2014})},\ \Eprint {http://arxiv.org/abs/1311.0815} {arXiv:1311.0815
  [hep-th]} \BibitemShut {NoStop}%
\bibitem [{\citenamefont {Lobo}\ \emph {et~al.}(2013)\citenamefont {Lobo},
  \citenamefont {Olmo},\ and\ \citenamefont {Rubiera-Garcia}}]{Lobo:2013prg}%
  \BibitemOpen
  \bibfield  {author} {\bibinfo {author} {\bibfnamefont {F.~S.~N.}\
  \bibnamefont {Lobo}}, \bibinfo {author} {\bibfnamefont {G.~J.}\ \bibnamefont
  {Olmo}}, \ and\ \bibinfo {author} {\bibfnamefont {D.}~\bibnamefont
  {Rubiera-Garcia}},\ }\href {\doibase 10.1088/1475-7516/2013/07/011}
  {\bibfield  {journal} {\bibinfo  {journal} {JCAP}\ }\textbf {\bibinfo
  {volume} {1307}},\ \bibinfo {pages} {011} (\bibinfo {year} {2013})},\ \Eprint
  {http://arxiv.org/abs/1306.2504} {arXiv:1306.2504 [hep-th]} \BibitemShut
  {NoStop}%
\bibitem [{\citenamefont {Chen}\ \emph {et~al.}(2015)\citenamefont {Chen},
  \citenamefont {Ong},\ and\ \citenamefont {Yeom}}]{Chen:2014jwq}%
  \BibitemOpen
  \bibfield  {author} {\bibinfo {author} {\bibfnamefont {P.}~\bibnamefont
  {Chen}}, \bibinfo {author} {\bibfnamefont {Y.~C.}\ \bibnamefont {Ong}}, \
  and\ \bibinfo {author} {\bibfnamefont {D.-h.}\ \bibnamefont {Yeom}},\ }\href
  {\doibase 10.1016/j.physrep.2015.10.007} {\bibfield  {journal} {\bibinfo
  {journal} {Phys. Rept.}\ }\textbf {\bibinfo {volume} {603}},\ \bibinfo
  {pages} {1} (\bibinfo {year} {2015})},\ \Eprint
  {http://arxiv.org/abs/1412.8366} {arXiv:1412.8366 [gr-qc]} \BibitemShut
  {NoStop}%
\end{thebibliography}%
  
\end{document}